\documentclass[11pt,leqno]{article}
\usepackage{graphicx}
\newtheorem{lemma}{Lemma} 
\newcommand{\nint}[1]{{\rm nint}\left(#1\right)}
 
\newcommand{\R}{\bf R} 
\newcommand{\Z}{\bf Z}

\newcommand{\ket}[1]{|#1\rangle}

\newcommand{\odd}{{\rm odd}}
\newcommand{\even}{{\rm even}}
\renewcommand{\k}{{\tilde k}}
\newcommand{\floor}[1]{\left\lfloor#1\right\rfloor}
\newcommand{\ceil}[1]{\left\lceil#1\right\rceil}
\newcommand{\nr}{\frac{\pi mj}{2^{n-\k}}}
\newcommand{\dr}{\frac{\pi j}{2^{n-\k}}}
\newcommand{\hsps}{\hspace{-.1in}}
\newcommand{\ha}{\hspace{-.15in}}
\newcommand{\vstrut}{\rule{0in}{0.15in}}
\newcommand{\Si}{{\rm Si}}
\newcommand{\vstrutt}{\rule{0in}{0.25in}}
\newcommand{\x}{x_{\rm m}}
\begin{document}
\begin{center} {\Large \sc Sharp probability estimates for Shor's
order-finding algorithm}
\bigskip

{\sc P.\ S.\ Bourdon}\footnote{Department of Mathematics, Washington and Lee University,
Lexington, VA 24450; pbourdon@wlu.edu} and {\sc H.\ T.\  Williams}\footnote{Department of Physics and Engineering, Washington and Lee
University, Lexington, VA 24450; williamsh@wlu.edu}

\end{center}

{\bf Abstract}: Let $N$ be a (large) positive integer, let $b$ be an integer satisfying $1< b < N$ that is
relatively prime to $N$, and let $r$ be the order of
$b$ modulo $N$.  Finally, let
 QC be a quantum computer whose input register has the size specified
in Shor's original description of his order-finding algorithm.  We
prove that when Shor's algorithm is implemented on QC, then the
probability $P$ of obtaining a (nontrivial) divisor of $r$ exceeds $.7$
whenever $N > 2^{11}$ and $r\ge 40$, and we establish that 
$.7736$ is an asymptotic lower bound for $P$.    When
$N$ is not a power of an odd prime, Gerjuoy has shown that $P$ exceeds
90 percent for $N$ and $r$ sufficiently large. We give easily checked
conditions on $N$ and $r$ for this 90 percent threshold
to hold,   and we establish an asymptotic lower bound for $P$ of
$2\Si(4\pi)/\pi \approx .9499$ in this situation.   More generally, for
any nonnegative integer $q$, we show that when QC$(q)$ is a quantum
computer whose input register has $q$ more qubits than does QC, and 
Shor's algorithm is run on QC$(q)$, then an asymptotic lower bound on
$P$ is $2\Si(2^{q+2}\pi)/\pi$ (if $N$ is not a power of an odd prime). 
Our arguments are elementary and our lower bounds on $P$ are carefully
justified.

   \section{Introduction}

  In this Introduction, we assume readers are familiar with Shor's
algorithm for finding the order of an integer $b$ relative to a larger
integer $N$ to which $b$ is relatively prime. The algorithm is reviewed in
the next section.  

   The goal of Shor's algorithm is to find the least positive integer $r$
such that $b^r \equiv 1\pmod{N}$; that is, to find the order of $b$ modulo
$N$.  In \cite{Shor1,Shor2}, Shor describes an efficient algorithm to
accomplish this task that runs on a quantum computer whose input register
has $n$ qubits, where $n$ is chosen to be the unique positive integer such
that $N^2 \le 2^n < 2N^2$. The final quantum-computational step in Shor's
algorithm is measurement of the input register in the computational basis. 
One obtains an $n$-bit integer $y$, and the key calculation at this point
is the probability  that $y$ satisfies
\begin{equation}\label{DE}
\left|y - \frac{s2^n}{r}\right| \le \frac{1}{2}\ \ {\rm for\ some}\ s\in
\{1, 2, \ldots r-1\}.
\end{equation}
  Lower bounds for this probability, for sufficiently large $N$  and $r$,
are typically given at around 40 percent along with
$4/\pi^2$ as  an asymptotic lower bound (see, e.g., \cite[p.\ 1500]{Shor2}, \cite{EkJo}, \cite[p.\
58]{Hir},  \cite[Chapter 3]{Mer}).    We find a precise formula for the
probability that  $y$ belongs to 
$$ S:=\left\{\nint{\frac{s2^n}{r}}: s= 1, 2, 3, \ldots, r-1\right\},
$$ and thereby satisfies (\ref{DE}).  Here, {\it nint} is the
nearest-integer function. We use our probability formula to show that the
integer
$y$ obtained by Shor's quantum computation will belong to $S$ with
probability exceeding  
$70\%$, as long as $N\ge 2^{11}$ and $r\ge 40$.  In fact, we show that the probability $P$ that $y$ belongs to $S$ will
exceed $70\%$ as long as $N\cdot2^{11} \le 2^n$ and $r\ge 40$.  Moreover, we show that 
$$
 \frac{2}{\pi^2}(-2+\pi \Si(\pi)) \approx 0.7737
$$
provides an asymptotic lower bound for $P$. 
Here $\Si$ is the  sine-integral function  $\Si(x) =
\int_0^x\frac{\sin{t}}{t}\, dt$.  
Note that we may assume both $r$ and
$N$ are large; otherwise, there is no reason to resort to quantum
computation to find $r$.

 Efficient order finding leads to efficient methods for factoring composite
integers (see, e.g., \cite[\S 5.3.2]{NC}).  
  Interest in the factoring problem is especially great for composite
integers of the form $pq$, where $p$ and $q$ large distinct primes---the
ability to factor such integers is equivalent to the ability to read
information encoded via the RSA cryptography system (see, e.g,
\cite{Ger}).   When $N$ is not a power of a prime,   Gerjuoy
(\cite{Ger}) shows that Shor's algorithm (input register having $n$ qubits
where $N^2\le 2^n < 2N^2$) succeeds in finding a divisor of $r$ with
probability  exceeding  90\%, given
$N$ and
$r$ are sufficiently large.  (Here and in the sequel, by ``divisor of $r$''
we mean a divisor exceeding $1$.) The key lemma for Gerjuoy's work is that
$r < N/2$ whenever $N$ is not a power of a prime. (See
\cite[Appendix B]{Ger} for an elementary proof of this fact in case
$N = pq$, where $p$ and $q$ are distinct odd  primes; we provide  a general
argument in Section~\ref{GerSect} below.)   This lemma allows Gerjuoy to establish that
Shor's algorithm finds a divisor of $r$ whenever the integer $y$ observed
at the conclusion of quantum computation belongs to 
$$
\tilde{S} : = \left\{y: \left|y - \frac{s2^n}{r}\right| \le 2\ \ {\rm for\
some}\ s\in \{1, 2, \ldots, r-1\}\right\}.
$$ In Section~\ref{GerSect} of this paper, we apply our methods to find a precise
formula for the probability $\tilde{P}$ that $y$ belongs to
$\tilde{S}$.  We then use the formula to describe conditions on $r$ and $N$
that will ensure $\tilde{P}$ exceeds 90\%
 and we show that 
$$
\frac{2\Si(4\pi)}{\pi} \approx 0.9499
$$
is an asymptotic lower bound for $\tilde{P}$.  

   In the final section of this paper, we extend our results to the case
where the quantum computer ``QC(q)'' implementing Shor's algorithm has $n +
q$ qubits in its input register, where  $q$ is a nonnegative integer and where,
just as before,
$N^2\le  2^n < 2N^2$.  Again assuming that $N$ is not a power of a prime
(so that Gerjuoy's lemma applies),  we show that when Shor's algorithm is
run on QC(q), an asymptotic lower bound on the probability of finding a
divisor of $r$ is
$$
\frac{2\Si(2^{2+q}\pi)}{\pi}.
$$ When $q = 3$, our asymptotic bound exceeds $0.993$. Also, when $q= 3$, we
give easily checked conditions on $r$ and $N$ that will ensure the
probability of success exceeds 99 percent. 

We remark that  phase-estimation analysis, as it is described in
 in \cite{NC} (see the paragraph containing (5.44) on page 227),
assures that the 99 percent threshold is reached when $q=5$  ($N$ not a
power of a prime), or when $q=7$ ($N$ arbitrary).

\section{Preliminaries}   Our probability analysis depends on some
elementary number theory; specifically, the following two lemmas.  In these
lemmas, $r$ is a positive integer exceeding $1$.

\begin{lemma}\label{RPL}  Suppose that $t$ is a positive integer less than
$r$ which is relatively prime to $r$ and that $k$ is a positive integer;
then $\{(kr +s)t\pmod{r}: s = 1, 2,
\ldots r-1\} = \{1, 2, \ldots, r-1\}$.
\end{lemma}

\begin{quote} {\it Proof}: Define $f: \{1, 2, \ldots, r-1\}\rightarrow \{1, 2,
\ldots, r-1\}$ by 
$$ f(s) = (kr + s)t\hsps\pmod{r} = st\hsps\pmod{r}.
$$
  To prove the lemma, it suffices to show $f$ is one-to-one.  Suppose
$f(s_1) = f(s_2)$, then 
$$ (s_1  - s_2) t \equiv 0\hsps\pmod{r}.
$$ Since $t$ is relatively prime to $r$, the preceding equation shows that
$r$ must divide $s_1 - s_2$, but since $|s_1 - s_2| < r$, we must have
$s_1 -s_2 = 0$.  Hence $f$ is one-to-one, as desired.///
\end{quote}

\begin{lemma}\label{KL} Suppose that $2^n$ exceeds $r$ and $r = 2^{\k}r'$,
where
$\k$ is a nonnegative integer and $r'$ is a positive odd integer exceeding
$1$.  Then there exists an integer $q$ and a positive integer
$t$ less than $r'$, relatively prime to $r'$, such that
$$
\frac{2^n}{r} = q + \frac{t}{r'}.
$$
\end{lemma}

\begin{quote}{\it Proof}:  Note $2^n/r = 2^{n-\k}/r'$. Let $q$ be the integer quotient
that results when $2^{n-\k}$ is divided by $r'$ and let $t$ be the
remainder:
$$
\frac{2^{n-\k}}{r'} = q + t/r'.
$$ It follows that $2^{n-\k} = q r' + t$, and this equation shows that if
$t$ and
$r'$ had a common divisor exceeding $1$ (necessarily odd since $r'$ is odd), then that common
divisor would be a odd number greater than $1$ dividing $2^{n-\k}$, which is absurd.  The
lemma follows.///
\end{quote}

 Let $\Z^+$ denote the set of positive integers.  For the remainder of this
paper, $b$ and $N$ denote elements of
$\Z^+$ such that
$b < N$ and $b$ is relatively prime to $N$.  Let
$r$ be the {\it order} of $b$ modulo $N$:  $b^r \equiv 1 \pmod{N}$ and $r$
is the least positive integer for which this equation holds.  It is easy to
show such an $r$ exists and that $r < N$.\footnote{In fact, $r$ must divide
$\phi(N)$, where $\phi(N)$ denotes the number of positive integers less
than $N$ that are relatively prime to $N$.} Also, since $1 < b < N$, $r$ must be greater than $1$.  We now describe Shor's quantum
algorithm, which is designed to compute the order
$r$ of $b$.  

We focus on the transformations and measurements of the input and output
registers of the machine implementing the algorithm, ignoring any
work-register activity.   The machine has input register having
$n$ qubits, where
$n$ is the least positive integer such that 
$$ N^2 \le 2^n < 2N^2.
$$ Its output register will have $n_0$ qubits, where $n_0$ is the least
positive integer for which $N\le 2^{n_0}$. (It's easy to check that either
$n=2n_0$ or $n = 2n_0 - 1$.) Note that the size of the output register
allows it to hold any of the $r$ integers in the set $\{b^x\hsps\pmod{N}: x = 0,
1, 2, \ldots, r-1\}$.

The machine begins in state $\ket{0}_n\ket{0}_{n_0}$. Then Hadamard gates
are applied to each of the $n$ qubits in the input register to put the machine in
state
$$
\frac{1}{2^{n/2}} \sum_{x=0}^{2^n-1} \ket{x}_n\ket{0}_{n_0}.
$$  Then the unitary transformation that takes
$$\ket{x}_n\ket{0}_{n_0}\ \ {\rm to} \ \
\ket{x}_n\ket{b^x\ha\pmod{N}}_{n_0}, x\in \{0, 1, 2, \ldots, 2^n-1\},
$$ is applied, yielding the machine state
\begin{equation}\label{FM}
\frac{1}{2^{n/2}} \sum_{x=0}^{2^n-1} \ket{x}_n\ket{b^x\ha\pmod{N}}_{n_0}.
\end{equation} The next step in the algorithm, as described by Shor
\cite{Shor2}, is the application of the quantum Fourier-transform to the
input register.  However, to limit the number of summations that appear in
our work, we will, at this stage, follow David Mermin \cite[Chapter 3]{Mer}
and measure the output register.  When this measurement is made on the
machine in state (\ref{FM}), we obtain an $n_0$-bit integer $J$. Observe
that there must be exactly one integer $x_0$ in $\{0, 1, 2, \ldots, r-1\}$
such that $b^{x_0} \equiv J\pmod{N}$ and that every
$x\in\{0,1, \ldots, 2^n-1\}$ such that $b^x \equiv J\pmod{N}$ has the form $x_0 +
kr$ for some integer $k$ in $\{0,1,2, \ldots,m-1\}$, where 
\begin{equation}\label{mdef} m =  \ceil{\frac{2^n}{r} - \frac{x_0}{r}}.
\end{equation} (Here, $\ceil{w}$ represents the least integer greater than
or equal to the real number $w$;  later we use $\floor{w}$ to represent
that greatest integer less than or equal to $w$.)  For future reference,
observe
\begin{equation}\label{mbounds}
 \frac{2^n}{r} - 1 < m <  \frac{2^n}{r} +1.
\end{equation}  Thus, after measuring $J$ in the output register, the
machine's input register is in state
\begin{equation}\label{BQFT}
\frac{1}{\sqrt{m}}\sum_{k=0}^{m-1}\ket{kr + x_0}_n.
\end{equation} We can think of the input register's state as $1/\sqrt{m}$
times  a vector of $0$'s and $1$'s, which has
$1$'s in positions
$kr + x_0$ for $k\in \{0, 1, 2, \ldots, m-1\}$ and zeros elsewhere.  Thus
the input register contains values of a periodic
$\{0, 1/\sqrt{m}\}$-valued function, having  period
$r$.   By taking the quantum Fourier-transform of the input register, we
(hope to) obtain information about the fundamental frequency $1/r$ and its
overtones $s/r$, $s = 2, 3, \ldots, r-1$.   After applying the quantum
Fourier-transform, the input register is in state
\begin{equation}\label{FPM}
\frac{1}{\sqrt{2^nm}}\sum_{y=0}^{2^n-1}e^{2\pi i x_0 y/2^n}\sum_{k=0}^{m-1}
e^{2\pi i kry/2^n}\ket{y}.
\end{equation} Here we are following Mermin \cite[Chapter 3]{Mer}, even
notationally.  

The final step in the quantum-computational part of Shor's algorithm is 
measurement of the input register, which yields and $n$-bit integer
$y$.   The key calculation at this point is the probability that the
integer $y\in \{0, 1, 2, \ldots, 2^n -1\}$ measured will belong to 
\begin{equation}\label{SetNI} S= \left\{\nint{\frac{s2^n}{r}}: s= 1, 2, 3,
\ldots, r-1\right\}.
\end{equation}

This calculation is carried out in the next section.

\section{An Exact Probability Calculation}\label{SectT}

For each $s\in \{1, 2, \ldots, r-1\}$, let
$$ y_s = \nint{\frac{s2^n}{r}}.
$$ We seek to compute
\begin{quotation}{\em 
$P: =$  the probability that the $n$-bit integer $y$ observed via measurement of the quantum system in state
(\ref{FPM}) belongs to $S =
\{y_s: s = 1, 2, \ldots, r-1\}$. }
\end{quotation}
  If $y$ does belong to $S$, then Shor \cite{Shor1,Shor2} explains how to use that information to find a divisor of
$r$ (in an efficient way). He depends on a classical result in number theory that states that if
$y$ is an integer such that
\begin{equation}\label{NTG}
\left|\frac{y}{2^n} - \frac{s}{r}\right| \le \frac{1}{2r^2} \ \ {\rm for\ some}\ s\in \{1, 2, \ldots, r-1\},
\end{equation} then one can obtain, via the continued-fraction expansion of $y/2^n$, a  rational number
$\frac{\tilde{s}}{\tilde{r}}$, in lowest terms, such that $\frac{\tilde{s}}{\tilde{r}} = \frac{s}{r}$; hence, $r =
\frac{s}{\tilde{s}} \tilde{r}$ and $\tilde{r}$ is a divisor of $r$. If $s$ happens to be relatively prime to $r$, then
the order $r$ is determined.  Note that if $y = y_{s}$ is an element  of $S$, then
$$
\left|y - \frac{s2^n}{r}\right| \le \frac{1}{2},
$$ so that $|y/2^n - s/r| \le \frac{1}{2\ 2^n} \le \frac{1}{2N^2} <
\frac{1}{2r^2}$.  Thus observing an integer from $S$ at the conclusion of quantum computation will yield a divisor of
$r$. The probability of finding
$r$ itself, as the least common-multiple of divisors found, rises quickly to $1$ with the number of different divisors
known.

It follows from (\ref{FPM}) that the probability $p(y_s)$ that $y_s$ will be observed is
$$ p(y_s) = \frac{1}{2^nm}\left|\sum_{k=0}^{m-1}e^{2\pi i k r y_s/2^n}\right|^2, s\in \{1, 2, \ldots, r-1\},
$$ which may be rewritten,
\begin{equation}\label{PYS} p(y_s) =
\frac{1}{2^nm}\left|\sum_{k=0}^{m-1}e^{\frac{2\pi i kr}{2^{n}}\left(y_s -
\frac{s2^n}{r}\right)}\right|^2, s\in
\{1, 2,
\ldots, r-1\},
\end{equation} because  $e^{\frac{2\pi i kr}{2^{n}}\left(y_s -
\frac{s2^n}{r}\right)} = e^{2\pi i k r y_s/2^n}e^{-2\pi i ks} = e^{2\pi i k r y_s/2^n}$ for each $s$.  Let 
\begin{equation}\label{deltadef}
\delta_s = y_s - \frac{s2^n}{r} = y_s - \frac{s2^{n-\k}}{r'},
\end{equation} which allows us to re-express (\ref{PYS}) as
\begin{equation}\label{pdelta} p(y_s) =
\frac{1}{2^nm}\left|\sum_{k=0}^{m-1}e^{\frac{2\pi i kr\delta_s}{2^{n}}}\right|^2, s\in
\{1, 2,
\ldots, r-1\}.
\end{equation} Representation (\ref{pdelta}) of $p(y_s)$ can be simplified by using the formula for the partial sum of
a geometric series ($\sum_{l=0}^{m-1} w^l = (1-w^m)/(1-w)$; one obtains that for every $s\in \{1, 2, \ldots, r'-1\}$
\begin{equation}\label{probwsine} p(y_s) = \frac{1}{2^nm}\frac{\left|1 - e^{\frac{2\pi i
mr\delta_s}{2^n}}\right|^2}{\left|1-e^{\frac{2\pi i r\delta_s}{2^n}}\right|^2}  =
\frac{1}{2^nm}\frac{\sin^2\left(\frac{\pi m r
\delta_s}{2^n}\right)}{\sin^2\left(\frac{\pi r \delta_s}{2^n}\right)}.
\end{equation}

  In our calculation of $P$, we will assume only that the number $n$ of qubits in the input register exceeds $n_0$,
the number in the output register.  Observe that this ensures that $2^n/r > 2^{n-n_0} \ge 2$.  It follows that the set
$S$ of (\ref{SetNI}) consists of $r-1$ distinct elements, and thus
\begin{equation} \label{Pdef} P = \sum_{s=1}^{r-1} p(y_s).
\end{equation} Also note that there is no ambiguity in the value of $\nint{s2^n/r}$ for $s= 1, 2, \ldots, r-1$ because
$s2^n/r$ can never be a half-integer.

We consider the simplest case first:  $\frac{2^n}{r}$ is an integer.   In this case, $y_s =
\nint{\frac{s 2^n}{r}} = \frac{s 2^n}{r}$ and therefore $\delta_s = 0$ for each $s$.  It follows from (\ref{pdelta})
that $p(y_s) = \frac{m}{2^n}$ for every $s$ and thus
\begin{equation} P = (r-1) \frac{m}{2^n}.
\end{equation} Using the lower bound on m from (\ref{mbounds}) it is easy to show
\begin{equation} P > 1 - \frac{1}{r} - \frac{r-1}{2^n}.
\end{equation} This exceeds .95 if, e.g., $r>25$ and $n>15$.

We now address the more challenging, more interesting case: $\frac{2^n}{r}$ is not an integer.    Note that in this
case,  there must be a nonnegative  integer
$\k$ such that
\begin{equation}\label{rrep} r=2^\k r',\ {\rm where}\ r'\ {\rm is\ odd\ and\ exceeds}\ 1.
\end{equation} Suppose that $\k$ is positive so that $r$ is even. Then appearing in the sum over
$s$ in (\ref{Pdef}) are values of $s$ that are multiples of $r'$: $r', 2r', ..., (2^{\k}-1)r'.$  For each of these
values, $y_s =
\nint{\frac{s 2^n}{r}} = \nint{2^{n-\k}\frac{s}{r'}} = 2^{n-\k}\frac{s}{r'} $ and therefore $\delta_s = 0$ for each
such $s$. Thus we have from (\ref{pdelta})
\begin{quotation} {\bf Observation 1:} The total contribution to $P$ from multiples of $r'$ is $(2^{\k}-1)
\frac{m}{2^n}$.
\end{quotation} The remaining  $s$ values, i.e. those that are not multiples of $r'$, consist of $2^{\k}$ sequences,
each with
$r'-1$ terms:
\begin{equation}\label{sseqs}
 (1,2,...,r'-1), (r'+1, r'+2,...,2r'-1),  \ldots, 
\left(\vstrut (2^{\k}-1)r'+1,(2^{\k}-1)+2,...,2^{\k}r'-1\right).
\end{equation} We will show that the contribution to P from each sequence is the same.

Note that Observation 1 is valid even if $\k =0$ and that the assertion made in the preceding paragraph is trivially
true since there is only one sequence in (\ref{sseqs}) in this case.

Apply Lemma~\ref{KL} to represent $\frac{2^n}{r}$ as $q + t/r'$ where $q$ is a positive integer and $t< r'$ is
relatively prime to $r'$.    Consider the collection 
\begin{eqnarray*} S':= \{y_s:  s\in \{1, 2, \ldots, r'-1\}\} & = &
\left\{\nint{\frac{s2^n}{r}}: s\in \{1, 2, \ldots, r'-1\}\right\}\\  & = &
\left\{\nint{sq + \frac{st}{r'}}: s\in \{1, 2, \ldots, r'-1\}\right\}
\end{eqnarray*} Apply Lemma~\ref{RPL} with $k =0$ to see that $st\not\equiv 0\pmod{r'}$ for each $s\in \{1, 2, \ldots,
r'-1\}$. Hence, for each such
$s$, 
$st = q_sr' + j_s$  for some nonnegative integer $q_s$ and some $j_s \in
\{1, 2, \ldots, r'-1\}$. Thus 

\begin{equation}\label{FL2} S' = \left\{ \nint{sq + q_s + \frac{j_s}{r'}}: s\in \{1, 2, \ldots, r'-1\}\right\}. 
\end{equation} Let $s\in \{1, 2, \ldots, r'-1\}$. Observe that if 
\begin{equation}\label{O1} j_s \le \floor{\frac{r'}{2}}, {\rm then}\ y_s = sq + q_s\  {\rm and}\ y_s - \frac{s2^n}{r}
= -j_s/r'.
\end{equation} If
\begin{equation}\label{O2} j_s\ge \ceil{\frac{r'}{2}}, \ {\rm then}\ y_s = sq + q_s + 1\ {\rm and}\  y_s -
\frac{s2^n}{r} =  \frac{r'-j_s}{r'}.
\end{equation}

Lemma~\ref{RPL} tells us that as $s$ varies from 1 to $r' -1$, the integers
$j_s$ appearing in the representation
$\frac{s2^n}{r} = sq +
\frac{st}{r'} = sq + q_s + \frac{j_s}{r'}$  will also vary from $1$ to
$r'-1$. Thus, in (\ref{FL2}) 
$$
\left\{\frac{j_s}{r'}: s\in \{1, 2, \ldots, r'-1\}\right\} =
\left\{\frac{1}{r'}, \frac{2}{r'}, \ldots,
\frac{r'-1}{r'}\right\}. 
$$ Thus  Lemma~\ref{RPL} (with $k = 0$), combined with observations (\ref{O1}) and (\ref{O2}), yields
\begin{equation}\label{koc}
\left\{\left|y_s -  \frac{s2^n}{r}\right|: s  = 1, 2, \ldots, r' -1\right\} = \left\{\frac{1}{r'},
\frac{2}{r'}, \dots, \frac{\floor{r'/2}}{r'}\right\}
\end{equation}
 and that for a given $l\in \{1, 2, \ldots \floor{\frac{r'}{2}}\}$, there are exactly two integers $s_1$ and $s_2$ in
$\{1, 2, 3, \ldots, r'-1\}$ such that $|y_{s_1} - \frac{s_12^n}{r}| = l/r'$ and $|y_{s_2} -
\frac{s_22^n}{r}| = l/r'$.  Now suppose that $r' < r$;  in other words, the integer $\k$ in (\ref{rrep}) is positive. 
Let $k$ be any integer satisfying $1\le k \le 2^{\k} -1$.   The analysis of the preceding two paragraphs, with
Lemma~\ref{KL} applied as stated, shows that
\begin{equation}\label{ysr}
\left\{\left|y_s -  \frac{s2^n}{r}\right|: s  = kr'+ 1, kr' + 2, \ldots, kr' + r' - 1\right\} =
\left\{\frac{1}{r'},
\frac{2}{r'}, \dots, \frac{\floor{r'/2}}{r'}\right\},
\end{equation} with each element of the set on the right corresponding to
$|y_s -  \frac{s2^n}{r}|$ for exactly two values of
$s$ in the range $kr' +1$ to $kr' + r' - 1$.

Using the definition of $\delta_s$ from (\ref{deltadef}) as well as (\ref{koc}) and (\ref{ysr}), we see that for any
$k$ with $0\le k \le 2^\k -1$,
\begin{equation}\label{iddelta}
\{|\delta_{kr' + q}|: q = 1, 2, \ldots, r'-1\} = \left\{\frac{j}{r'}: j =1, 2, \ldots, \floor{\frac{r'}{2}}\right\}
\end{equation}
 with each member of the set on the right corresponding to $|\delta_{kr' + q}|$ for exactly two values of $q\in\{1,
2, \ldots, r'-1\}$. Thus we have
\begin{quotation}{\bf Observation 2:}  the contribution to $P$ from from any one of the sequences in (\ref{sseqs}),
which would take the form $\sum_{q=1}^{r'-1} p(y_{kr' +q})$ for some $k\in\{0, 1, \ldots, 2^\k-1\}$, is given by
\begin{equation}\label{FABY}
\frac{2}{2^nm}\sum_{j=1}^{\floor{\frac{r'}{2}}}\frac{\sin^2\left(\frac{\pi m r
(j/r')}{2^n}\right)}{\sin^2\left(\frac{\pi r (j/r')}{2^n}\right)} =
\frac{2}{2^nm}\sum_{j=1}^{\floor{\frac{r'}{2}}}\frac{\sin^2\left(\frac{\pi m j
}{2^{n-\k}}\right)}{\sin^2\left(\frac{\pi j}{2^{n-\k}}\right)},
\end{equation}
 where we have used (\ref{probwsine}).  
\end{quotation}

Combining Observations 1 and 2 leads us to a final form for the exact probability:
\begin{equation} P = 2^{\k}\frac{2}{2^nm}\sum_{j=1}^{\floor{r'/2}}\frac{\sin^2\left(\frac{\pi
mj}{2^{n-\k}}\right)}{\sin^2\left(\frac{\pi j}{2^{n-\k}}\right)}+ (2^{\k}-1)\frac{m}{2^n}.\label{Pform}
\end{equation} Note that the preceding formula is valid even when $\frac{2^n}{r}$ is an integer, provided we take $r =
2^\k r'$, where $r' =1$, and we follow convention and interpret the sum from $j =1$ to $j = \floor{\frac{r'}{2}} = 0$
to be $0$.

\section{Lower Bounds on the Probability of Success}\label{LowerBS}

In this section, we discuss two different ways of obtaining lower bounds for
$$ 
P =
2^{\k}\frac{2}{2^nm}\sum_{j=1}^{\floor{r'/2}}\frac{\sin^2\left(\frac{\pi
mj}{2^{n-\k}}\right)}{\sin^2\left(\frac{\pi j}{2^{n-\k}}\right)}+
(2^{\k}-1)\frac{m}{2^n},
$$
where $r < N\le 2^{n_0}$, $2^\k r' = r$ with $r'\ge 3$  odd (and $\k\ge 0$), and
$\frac{2^n}{r} - 1 < m < \frac{2^n}{r} +1$.  Our first method of bounding
$P$ below uses elementary inequalities based on the Maclaurin series for the sine function and requires only that $n> n_0$. Our
second method provides an integral-based underestimate and requires $N^2\le
2^n$ (Shor's condition).  The lower bounds presented below are
rigorously justified in Appendix B.

  To derive a series-based lower bound for $P$, we use the following
elementary inequalities:
\begin{equation}\label{sineEst}
\sin^2x \le x^2 \ {\rm for\ all}\ x,\ \ {\rm and} \ \   \sin^2 x\ge \left(x - \frac{x^3}{6}\right)^2\ {\rm for,\ say,}\ x\in \left[0,
\frac{3\pi}{4}\right].
\end{equation} 

We obtain (see Appendix B)
\begin{equation}\label{PLOD}
P>  \left(1- \frac{1}{2^{n-n_0}} -
\frac{1}{r}\right)\left(1 -
\frac{\pi^2}{36}\left(\frac{r+1}{r} + \frac{1}{2^{n -n_0 - 1}} +
\frac{1}{2^{2(n-n_0)}}\right)
\right)\ \ {\rm if}\ \k = 0\ (r, {\rm  odd}),
\end{equation}
and
\begin{eqnarray}\label{PLEVE}
P &>& \left(1- \frac{1}{2^{n-n_0}} - \frac{1}{r'}\right)\left(1 -
\frac{\pi^2}{36}\left(\frac{r'+1}{r'} + \frac{1}{2^{n-n_0-2}} +
\frac{1}{2^{2(n-n_0)-1}}\right)
\right)\\
& & \rule{1in}{0in} + \frac{1}{r'}- \frac{1}{2^\k r'} -
\frac{1}{2^{n-n_0}}\ \  {\rm if} \  \k > 0\ (r, {\rm even}).\nonumber
\end{eqnarray}

Assuming that $n-n_0 \ge 11$ and $r\ge 40$, one can show (Appendix B)
that the right-hand side of either (\ref{PLOD}) or (\ref{PLEVE}) exceeds
$0.70$.  Thus if Shor's algorithm is carried out with an input register having the
size described in Shor's original paper, then the probability of finding a
divisor of the period sought exceeds 70\% (as long as $r\ge 40$ and $N\ge
2^{11}$).

  Note that as $r$ and $n-n_0$ approach $\infty$ in our lower bound
formula (\ref{PLOD}) for odd $r$, we get an asymptotic lower bound on
$P$ of
$(1-\pi^2/36)\approx 0.726$.  A sharper asymptotic bound is provided by
\begin{equation}\label{PFinal}
P \ge \frac{1-\frac{\pi^2}{4N^2}}{1 + \frac{1}{N}}\left(\frac{2}{\pi^2}\int_{1/r'}^{\frac{1}{2} + \frac{1}{2r'}}\frac{\sin^2(\pi x)}{x^2}\, dx\right) -
\frac{3}{N} +\frac{1}{r'} -\frac{1}{2^\k r'},
\end{equation}
an inequality proved to be valid in Appendix B (assuming $N^2\le 2^n$). By letting $r'$ and $N$ approach infinity, we obtain 
$$
\frac{2}{\pi^2}\int_{0}^{1/2}\frac{\sin^2(\pi x)}{x^2}\, dx =   \frac{2}{\pi^2}(-2+\pi Si(\pi)) \approx 0.7737
$$
 as an asymptotic lower bound for $P$.

 Consider the function $F(N, \k, r')$ defined by the right-hand side of (\ref{PFinal}).  It is clear that if either of $\k$ or $N$ increases, so
does
$F$. Additionally, the partial derivative of $F$  with respect to $r'$ is positive (whenever $N$ exceeds, say $9$) and
thus $F$ increases in $r'$ as  well.  $F$ exceeds 0.75 when
$N = 2^{11}, r' = 75$, and $\k = 0$.  Thus if one uses a classical computer to check that the order
$r$ of $b$ modulo
$N$ doesn't have the form $2^\k c$ where $c$ is an odd number satisfying $1\le c\le 73$, and $\k$ is a nonnegative integer for which $2^\k c < N$,
then one can be over 75\% certain of success.  Note there are fewer than
$37
\log_2(N)$ numbers to check so that the checking may be done efficiently on a classical computer.  The 0.77 success-rate threshold is reached by, e.g.
$N= 2^{15}$, $r' = 447$.

\section{Order-finding when $N$ is not a power of a prime}\label{GerSect}

 In this section, we assume that $N$ is  not a power of prime, that $b$ is an integer satisfying $1 < b < N$ which is relatively prime to $N$, and that
$r$ is the order of $b$ modulo $N$.    In this situation,   Gerjuoy (\cite{Ger}) has
shown that Shor's algorithm succeeds in finding a divisor of $r$ with probability
on the order of $90\%$ (given that $N$ and $r$ are sufficiently large).  As we
mentioned in the Introduction, the key to Gerjuoy's work is his use of the
following lemma:

\begin{lemma}[Gerjuoy]\label{orderLem} If $N$ is not a power of a prime and $b$ is relatively prime to $N$, then the order $r$ of $b$ modulo $N$
must satisfy
$$
r < \frac{N}{2}.
$$
\end{lemma}

Gerjuoy \cite[Appendix B]{Ger} provides an elementary proof of the preceding lemma
in case $N = pq$, where $p$ and $q$ are distinct odd primes.  The general result
may be established as follows.  The collection of all integers less than $N$ and
relatively prime to $N$ forms a group under multiplication modulo $N$.  This
group, frequently denoted $U(N)$, contains $\phi(N)$ elements, where $\phi$ is
the Euler $\phi$ function. A well-known number-theory result (see, e.g.,
\cite[Proposition 4.1.3]{IR}) shows that $U(N)$ contains an element having order
$\phi(N)$ modulo $N$ if and only if
$N$ is
$2$ or
$4$ or has the form
$p^j$ or $2p^j$, where $p$ is an odd prime and $j\in \Z^+$.  Thus if $N$ is a not a power of a prime, then either
\begin{itemize}
\item[(i)] $U(N)$ contains no element of order
$\phi(N)$, or
\item[(ii)] $N = 2p^j$ for some positive integer $j$ and some odd prime $p$.
\end{itemize}
Suppose that (i) holds, that $b\in U(N)$, and that $b$ has order $r$ modulo $N$.  Since the order $r$ of $b$ must divide the number of elements in
$U(N)$ (\cite[p.\ 43]{Her}) and since $r \ne \phi(N)$, we must have $\phi(N) = kr$ for some integer $k\ge 2$. Hence 
$$
r = \phi(N)/k \le \phi(N)/2 < N/2,
$$
as desired.  Suppose that $(ii)$ holds.  Then $U(N)$ does contain elements of order $\phi(N)$; however, an easy calculation shows $\phi(2p^j) = p^{j}
-  p^{j-1}$, which is less than $N/2$. Thus in case (ii)  holds, all elements of $U(N)$ have order less than $N/2$, which completes the proof of the
lemma.

  Gerjuoy \cite{Ger} explains how Lemma~\ref{orderLem} shows that a divisor of
$r$ may be extracted from the integer $y$ observed at the conclusion of Shor's
quantum computation {\em for a larger collection of $y$'s than those contained in
the set $S$ of integers nearest $s2^n/r$, $s=1, 2,\ldots , r-1$}.  Specifically,
he shows that if one observes an integer $y$ satisfying
\begin{equation}\label{DEG}
\left|y - \frac{s2^n}{r}\right| \le 2\ \ {\rm for\ some}\ s\in \{1, 2, \ldots
r-1\},
\end{equation}
then one can obtain a divisor of $r$.  To see why this is so, recall from
(\ref{NTG}) that the real goal of the computation is to find an integer $y$ satisfying
\begin{equation}\label{Sbig}
\left|\frac{y}{2^n} - \frac{s}{r}\right| \le \frac{1}{2r^2} \ \ {\rm for\ some}\ s\in \{1, 2, \ldots, r-1\}.
\end{equation}
Note that if (\ref{DEG}) holds and Lemma~\ref{orderLem} applies (so that $2r < N$), then 
$$
\left|\frac{y}{2^n} - \frac{s}{r}\right| \le  \frac{2}{2^n}\le \frac{2}{N^2} < \frac{2}{(2r)^2} = \frac{1}{2r^2},
$$
so that knowledge of $y$ means knowledge of a divisor of $r$.  

Thus,  given that $N$ is not a prime power, Gerjuoy establishes that Shor's computation
is successful provided the integer observed belongs to
$$
\tilde{S} = \left\{y: \left|y- \frac{s2^n}{r}\right| \le 2: s= 1, 2, \ldots, r-1\right\}.
$$
Observe that the gap between successive values of $s2^n/r$ exceeds $2^n/r\ge N^2/(N/2) = 2N$ so that the set of integers satisfying 
$\left|y- \frac{s2^n}{r}\right| \le 2$ will be disjoint from those satisfying $\left|y- \frac{s'2^n}{r}\right| \le 2$, given $s'\ne s$.

We now describe the elements of $\tilde{S}$ relative to the nearest integers $y_s$ (introduced earlier) and calculate the exact probability that the
integer
$y$ observed at the end of the Shor computation will belong to $\tilde{S}$.

Recall that for each $s\in \{1, 2, \ldots, r-1\}$,  $y_s = \nint{s2^n/r}$.  Note that if $y_s< s2^n/r$, then $\tilde{S}$ will contain, in addition to
$y_s$, the integers $y_s +1$, $y_s + 2$, and $y_s - 1$.  Similarly, if $y_s > s2^n/r$, then $\tilde{S}$ will contain $y_s, y_s + 1$, $y_s-1$, and
$y_s-2$. Finally, if $s2^n/r$ is an integer (so that in the notation of Section~\ref{SectT}, $s = kr'$
for some
$k$ satisfying $1\le k \le 2^{\k}-1$), then
$\tilde{S}$ will contain $y_s-2, y_s-1, y_s, y_s + 1, y_s + 2$.   We have computed the probability 
$P$ that the integer observed belongs to set
$\{y_s: s = 1, 2,
\ldots, r-1\}$, where
$y_s =
\nint{s2^n/r}$.  Similar methods will allow us to compute the probability that integers of the form $y_s + h$, $h\in \{-2, -1, 1, 2\}$ will be
observed.    In fact,  we compute the probability the $y_s + h$ is observed for
any integer $h$, but in this section will focus only the $|h|\le 2$ case.

Let $h\in \{-2, -1, 1, 2\}$ and $s\in \{1, 2, \ldots, r-1\}$ be arbitrary.  Substituting $y_s + h$ for $y_s$ in (\ref{PYS}) and using the definition of
$\delta_s$ in (\ref{deltadef}), we obtain the probability $p(y_s + h)$ that $y_s + h$ will be observed:
\begin{equation}\label{pyhform}
p(y_s + h) =  \frac{1}{2^nm}\left|\sum_{k=0}^{m-1}e^{\frac{2\pi i kr(h+\delta_s)}{2^{n}}}\right|^2 = \frac{1}{2^nm}\frac{\sin^2\left(\frac{\pi m r
(h +\delta_s)}{2^n}\right)}{\sin^2\left(\frac{\pi r (h +\delta_s)}{2^n}\right)}, s\in
\{1, 2,
\ldots, r-1\},
\end{equation}
which should be compared to (\ref{pdelta}) and (\ref{probwsine}).  Let $P_h = \sum_{s=1}^{r-1}p(y_s + h)$.   We compute $P_h$ just as we did $P$:
\begin{eqnarray*}
P_h & = & \sum_{s=1}^{r-1}p(y_s + h)\\
 & = & \sum_{k=0}^{2^{\k}-1}\left(\sum_{q=1}^{r'-1} p(y_{\vstrut kr' +
q} +h)\right) + \sum_{k=1}^{2^{\k}-1}p(y_{\vstrut kr'}+h)\nonumber\\
&=&
\sum_{k=0}^{2^{\k}-1}\left(\sum_{q=1}^{r'-1}\frac{1}{2^nm}\frac{\sin^2\left(\frac{\pi m r
(h + \delta_{kr' + q})}{2^n}\right)}{\sin^2\left(\frac{\pi r (h +\delta_{kr' + q})}{2^n}\right)}\right)+
\frac{2^{\k}-1}{2^nm}\frac{\sin^2\left(\frac{\pi m r
h}{2^n}\right)}{\sin^2\left(\frac{\pi r h}{2^n}\right)},\\
\end{eqnarray*}
where we have used (\ref{pyhform}) to obtain the final equality above. 
Recall from (\ref{O1}), (\ref{O2}), and (\ref{iddelta}) that $\{\delta_{kr' + q}: q = 1, 2, \ldots r'-1\} = \{j/r': j = 1, 2, \dots, \floor{r'/2}\} \cup
\{-j/r': j = 1, 2, \dots, \floor{r'/2}\}$. Thus, we can say
\begin{eqnarray*}
P_h & = & \sum_{k=0}^{2^{\k}-1}\frac{1}{2^nm}\left(\sum_{j=1}^{\floor{r'/2}}\frac{\sin^2\left(\frac{\pi
m r (h + j/r')}{2^n}\right)}{\sin^2\left(\frac{\pi r
(h + j/r')}{2^n}\right)} + \sum_{j=1}^{\floor{r'/2}}\frac{\sin^2\left(\frac{\pi
m r (h - j/r')}{2^n}\right)}{\sin^2\left(\frac{\pi r
(h - j/r')}{2^n}\right)}\right) +\frac{2^{\k}-1}{2^nm}\frac{\sin^2\left(\frac{\pi m r
h}{2^n}\right)}{\sin^2\left(\frac{\pi r h}{2^n}\right)}
\end{eqnarray*}

Observe that $P_h$ is an even function of $h$, i.e., $P_h = P_{-h}$. Thus we can say that the probability of observing an integer in $\tilde{S}$ is 
\begin{equation}\label{PofS}
P + 2P_{1} + Pt,
\end{equation}
where  $Pt$ is the probability that the following integers are observed: (a) $y_s + 2$, given $s2^N/r> y_s$, or (b) $y_s -2$,
given
$s2^n/r < y_s$, or (c) both
$y_s + 2$ and
$y_s - 2$, given $s2^n/r$ is an integer.   We have
$$
Pt = 2\sum_{k=0}^{2^{\k}-1}\frac{1}{2^nm}\left(\sum_{j=1}^{\floor{r'/2}}\frac{\sin^2\left(\frac{\pi
m r (2 - j/r')}{2^n}\right)}{\sin^2\left(\frac{\pi r
(2 - j/r')}{2^n}\right)}\right) +2\frac{2^{\k}-1}{2^nm}\frac{\sin^2\left(\frac{\pi m r
2}{2^n}\right)}{\sin^2\left(\frac{\pi r 2}{2^n}\right)}.
$$
Using our formulas for $P$, $P_1$, and $Pt$, and doing a bit of rearranging, we obtain the following as the probability that an element of $\tilde{S}$
will be observed:
\begin{eqnarray}\label{PtildeC}
\tilde{P} & =& 2^{\k}\frac{2}{2^nm}\sum_{h=-2}^1\sum_{j=1}^{\floor{r'/2}}\frac{\sin^2\left(\frac{\pi
mr(j/r'+h)}{2^{n}}\right)}{\sin^2\left(\frac{\pi r (j/r'+h)}{2^{n}}\right)} + (2^{\k}-1)\frac{m}{2^n}\\  
&  & \rule{0.5in}{0in} +
2\frac{2^{\k}-1}{2^nm}\frac{\sin^2\left(\frac{\pi m r 1}{2^n}\right)}{\sin^2\left(\frac{\pi r 1}{2^n}\right)}
+2\frac{2^{\k}-1}{2^nm}\frac{\sin^2\left(\frac{\pi m r 2}{2^n}\right)}{\sin^2\left(\frac{\pi r
2}{2^n}\right)}\nonumber
\end{eqnarray}

In Appendix A, we present a numerical calculation illustrating the correctness of our formula for $\tilde{P}$.
In Appendix B, we obtain the following lower bound for $\tilde{P}$:
\begin{eqnarray}
\tilde{P} &\ge&  \frac{\left(1 -
\frac{\pi^2}{N^2}\right)\left(1-\frac{\pi^2}{16N^2}\right)}{1 +
\frac{1}{2N}}\sum_{h=-2}^{1}\left( \frac{2}{\pi^2}\int_{0}^{\frac{1}{2} -
\frac{1}{2r'}}\frac{\sin^2(\pi x)}{(x+h)^2}\, dx\right)  - \frac{1}{r'}\label{NNPP}\\
& &\rule{2.5in}{0in} -\frac{7}{2N} -\frac{16}{\pi N\left(1-
\frac{1}{2N}\right)}  - \frac{1}{2^\k r'}.\nonumber
\end{eqnarray}
As $r'$ and $N$ approach infinity, we obtain an asymptotic lower bound of
\begin{equation}\label{LBG}
\sum_{h=-2}^1\left(\frac{2}{\pi^2}\int_{0}^{\frac{1}{2}}\frac{\sin^2(\pi x)}{(x +h)^2}\, dx\right) =
\frac{2\Si(4\pi)}{\pi} \approx 0.9499.
\end{equation}
 Clearly, the quantity on the right-hand side of (\ref{NNPP}) increases as any one of $r'$, $N$, or $\k$ increases. This quantity exceeds 0.90 when $N =
2^{16}, r' = 59$, and $\k = 0$.  Thus if one uses a classical computer to check that the order $r$ of $b$ modulo $N$ doesn't have the form $2^\k c$
where $c$ is an odd number satisfying $1\le c\le 57$, and
$\k$ is a nonnegative integer for which $2^\k c < N$, then one can be over 90\% certain of success.  Note there are fewer
than
$29
\log_2(N)$ numbers to check so that the checking may be done efficiently on a classical computer.  The 0.94 success-rate
threshold is reached by, e.g.
$N= 2^{16}$, $r' = 299$.

We remark that the trig identity $\sin^2(\pi x) = \sin^2(\pi(x + h))$ (for $h$, an integer) along with some elementary calculus shows that the sum of
integrals on the left of (\ref{LBG}) equals 
$$
\frac{2}{\pi^2}\int_0^2\frac{\sin^2(\pi x)}{x^2}\, dx
$$
which via appropriate trig identities and substitutions yields $\frac{2\Si(4\pi)}{\pi}$.

\section{Probability Calculations for Larger Computers}\label{PCLCSect}

 Just as in the preceding section, we assume that $N$ is a (large) positive integer that is not a power of a prime, that $b>1$ is an integer
than $N$, relatively prime to $N$, whose order $r$ (modulo $N$) we seek.  Note that Gerjuoy's lemma remains in force: $r < N/2$.  Just as before,
let $n$ be the positive integer satisfying $N^2\le 2^n < 2N^2$ so that $n$ is the number of qubits Shor originally specified for the input register
of the quantum computer ``QC'' running his order-finding algorithm.  For each nonnegative integer $q$ let QC(q) be a quantum computer having input
register of size $n + q$ qubits.  Let
$$
\tilde{S}_q = \left\{y: \left|y - \frac{s2^{n+q}}{r}\right| \le 2^{1 + q} \ {\rm for\ some}\ s\in \{1, 2, \ldots, r-1\} \right\}.
$$
Observe that if $y\in \tilde{S}_q$ then for some $s\in \{1, 2, \ldots, r-1\}$,
$$
\left|\frac{y}{2^{n+q}} - \frac{s}{r}\right| \le \frac{2}{2^n} \le \frac{2}{N^2} < \frac{2}{(2r)^2} =
\frac{1}{2r^2}
$$
so that if the integer $y$, observed at the end of the Shor computation on QC(q), belongs to $\tilde{S}_q$ then the computation will be
successful in the sense that a divisor of $r$ (exceeding $1$) will be found.  

   Let $\tilde{P}_q$ be the probability that a integer $y$ in $\tilde{S}_q$ is observed. (Note $\tilde{S}_0 = \tilde{S}$ and $\tilde{P}_0 =
\tilde{P}$.)  Generalizing the computations in the preceding section in the obvious way, we obtain the following analogue of (\ref{PtildeC}):
\begin{eqnarray}\label{PqC}
\tilde{P}_q & =&
2^{\k}\frac{2}{2^{n+q}m}\sum_{h=-2^{q+1}}^{2^{q+1}-1}\sum_{j=1}^{\floor{r'/2}}\frac{\sin^2\left(\frac{\pi
mr(j/r'+h)}{2^{n+q}}\right)}{\sin^2\left(\frac{\pi r (j/r'+h)}{2^{n+q}}\right)} +
(2^{\k}-1)\frac{m}{2^{n+q}}+ 
2\frac{2^{\k}-1}{2^{n+q}m}\sum_{j=1}^{2^{q+1}}\frac{\sin^2\left(\frac{\pi m r
j}{2^{n+q}}\right)}{\sin^2\left(\frac{\pi r j}{2^{n+q}}\right)}.\nonumber
\end{eqnarray}

The preceding formula yields  the following lower-bound for $\tilde{P}_q$ (see Appendix B), which is a  generalization of our lower-bound
formula  (\ref{NNPP})  for
$\tilde{P}$ :
\begin{eqnarray}
\tilde{P}_q &\ge&  \frac{\left(1 -
\left(\frac{\pi}{N}\right)^2\right)\left(1-\left(\frac{\pi}{2^{q+2}N}\right)^2\right)}{1 +
\frac{1}{2^{q+1}N}}\sum_{h=-2^{q+1}}^{2^{q+1}-1}\left( \frac{2}{\pi^2}\int_{0}^{\frac{1}{2} -
\frac{1}{2r'}}\frac{\sin^2(\pi x)}{(x+h)^2}\, dx\right)  - \frac{1}{r'}\label{GCPP}\\
& &\rule{2.4in}{0in}  - \frac{7}{N2^{q+1}} -\frac{16}{\pi N\left(1-
\frac{1}{N2^{q+1}}\right)}  - \frac{1}{2^\k r'}.\nonumber
\end{eqnarray}
Fixing $q$ and letting $N$ and $r'$ approach infinity, we obtain
$$
\sum_{h=-2^{q+1}}^{2^{q+1}-1}\left(\frac{2}{\pi^2}\int_{0}^{\frac{1}{2}}\frac{\sin^2(\pi x)}{(x +h)^2}\,
dx\right) =
\frac{2\Si(2^{2+q}\pi)}{\pi}
$$
as an asymptotic lower bound for $\tilde{P}_q$.  When $q = 3$, we have $\frac{2\Si(32\pi)}{\pi} \approx 0.9937$

Fix $q = 3$.  Clearly, the quantity on the right-hand side of (\ref{GCPP}) increases as any one of $r', N$, and $\k$ increases.  Given $q = 3$, this
quantity exceeds 0.99 when $N = 2^{20}, r' = 819$, and $\k = 0$.  Thus if one uses a classical computer to check that the order $r$ of $b$ modulo $N$
doesn't have the form $2^\k c$ where
$c$ is an odd number satisfying $1\le c\le 817$, and
$\k$ is a nonnegative integer for which $2^\k c < N$, then one can be over 99\% certain of success in
finding a divisor of $r$ on QC(3).  Note there are fewer than
$409
\log_2(N)$ numbers to check so that the checking may be done efficiently on a classical computer.

  Recall the well-known result 
$$
\Si(\infty) = \int_0^\infty \frac{\sin t}{t}\, dt = \frac{\pi}{2};
$$
thus, given Shor's algorithm runs on QC(q), our asymptotic lower bound $\frac{2\Si(2^{2+q}\pi)}{\pi}$ on the probability of success
approaches $1$ as $q\rightarrow \infty$, as expected.
\clearpage

\begin{flushleft}
{\Large \bf Appendix A: Some Numerical Calculations}
\end{flushleft}
\bigskip

To illustrate the correctness of our formula (\ref{PtildeC}) for $\tilde{P}$,  we
complete a case study here involving small values of $N$ and $r$: we take $N = 247$ and $b = 4$ so that $r = 18$, which means
$\k = 1$ and $r' = 9$.  We use {\it Maple} to calculate
$\tilde{P}$ two ways.
\begin{itemize}
\item[(1)]  We use the (inverse) discrete Fourier transform\footnote{As an
operator, the inverse of the discrete Fourier transform is equivalent to
what is called the quantum Fourier transform.} to compute the coordinates,
relative to the computational basis,  of the state  (\ref{FPM}), which is
the state  that results from applying the quantum Fourier transform to the
periodic vector (\ref{BQFT}).  We plot the resulting probability amplitudes
and sum those corresponding to basis states belonging to 
$$
\tilde{S} =\left\{y: \left|y - \frac{s2^n}{r}\right| \le 2\ \ {\rm for\
some}\ s\in \{1, 2, \ldots, r-1\}\right\}.
$$  
\item[(2)] We use our formula  (\ref{PtildeC}).
\end{itemize} The reader will see that the probabilities calculated by (1)
and (2) agree to many decimal places.
\bigskip

{\bf Maple Probability Calculation Based on Fourier Coefficients}
\smallskip

 We suppose $N= 247$ and  $b= 4$ so that $r= 18$. Here, the 
output register will have $n_0=8$ qubits and,  following Shor, the input 
register will have $n = 16$ qubits.  For simplicity we take $x_0 = 0$ in 
(\ref{BQFT}) and create a vector $V$ corresponding to this state. Then we 
apply {\tt InverseFourierTransform(V)}, plot the resulting probability amplitudes
and sum those corresponding to the possible desired outcomes---those in the $\tilde{S}$. 
 Here's the {\it
Maple} code and output.
\bigskip

$>$Digits:=20:

$>$with(DiscreteTransforms):

$>$V:=Vector(2\symbol{94}(16)): \# V will store values of
periodic function to which DFT applied; entries initialized to 0

$>$m:=ceil(2\symbol{94}(16)/18);

{\it m := 3641}

$>$for k from 0 to m-1 do V[k*18+1]:=1/sqrt(m): od:
\#Every 18th value of V set to 1/sqrt(m)

$>$Z:=InverseFourierTransform(V):

$>$ for k from 0 to 2\symbol{94}16-1 do NZ[k]:=Z[k+1] od: \#Re-index
 so that NZ[k] is amplitude of  $|k>$ for k = 0..$2^{16}$-1

$>$with(plots):

$>$pointplot({seq([p/2\symbol{94}(16),abs(NZ[p])\symbol{94}2],p=0..2\symbol{94}16-1)});

\begin{center}

\includegraphics*[height=2.5in]{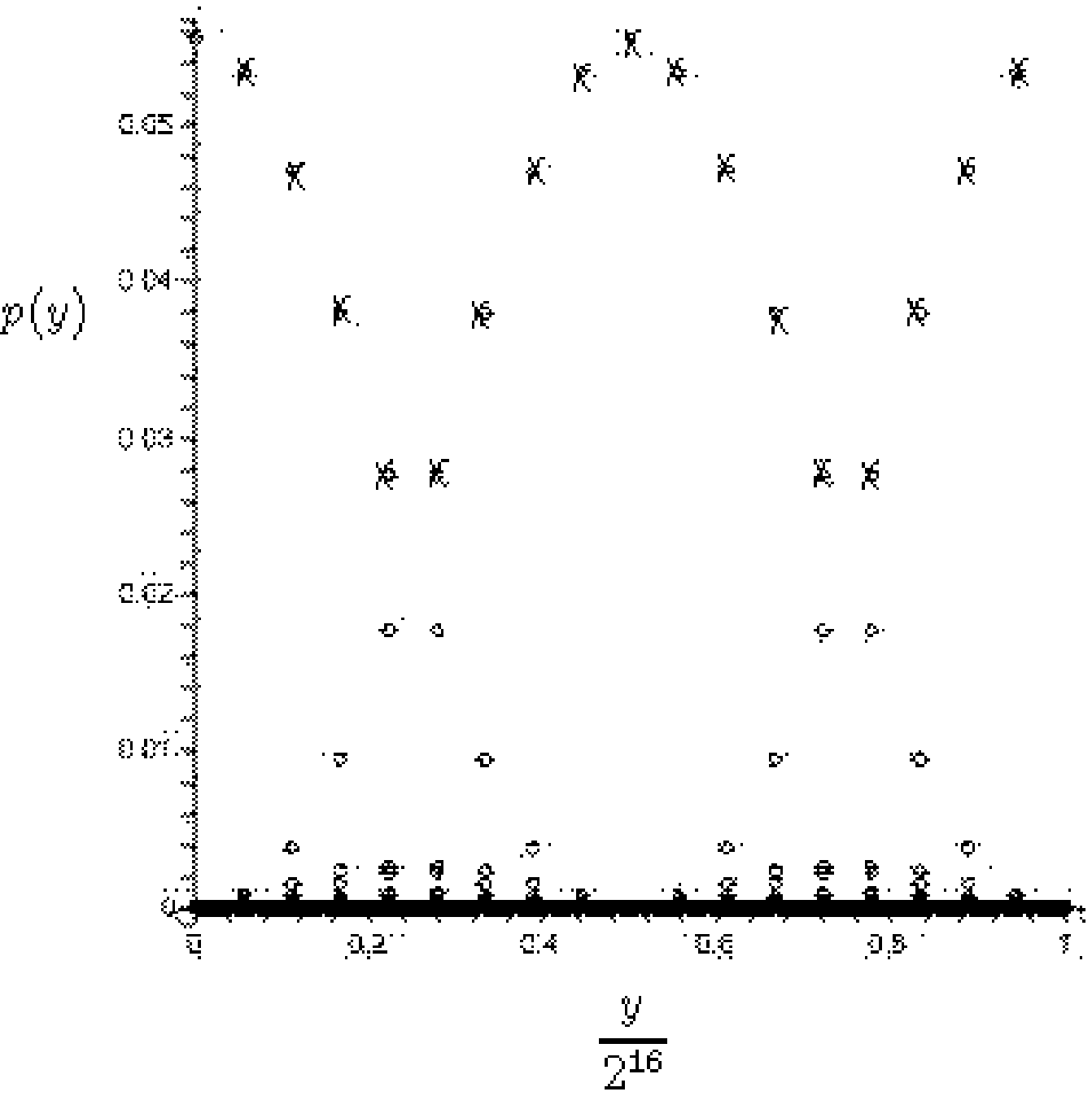}

\begin{minipage}{5in}
{\bf Figure 1}: The probability that  the integer $y$ is observed peaks when
$\frac{y}{2^{16}}$ is near an element of $\left\{\frac{s}{18}: s
=0,1,\ldots, 17\right\}$.  Data-points marked with x's have have coordinates $\left(\frac{y_s}{2^{16}}, p(y_s)\right)$, where $y_s$ is the integer
nearest $s2^{16}/18$, $s= 1, 2, \dots, 17$.
\end{minipage}

\end{center}

$>$for s from 1 to 17 do
y[s]:=round(s*2\symbol{94}(16)/18): od: \#Compute the nearest
integers

$>$Prob:=0: \#After next loop Prob will be probability of
observing an integer in $\{y[s]: s = 1..17\}$

$>$for k from 1 to 17 do
  Prob:= Prob + abs(NZ[y[k]])\symbol{94}2: od:

$>$ Prob;

.71982482558080545540

$>$Prob1:=0: \#After next two loops Prob1 will be probability of
observing an integer in $\{y[s]\pm1: s = 1..17\}$

$>$for k from 1 to 17 do Prob1:=Prob1 + abs(NZ[y[k]+1])\symbol{94}2: od:

$>$for k from 1 to 17 do Prob1:=Prob1 + abs(NZ[y[k]-1])\symbol{94}2: od:

$>$ Prob1;

                     0.15577667957639559817

$>$Prob2:=0: \#After next loop Prob2 will be probability of
observing the integer  y[s] + 2 or y[s] - 2, whichever is closer to s2\symbol{94}16/18

$>$ for k from 1 to 17 do
 
if (round(k*2\symbol{94}(16)/18) $<$ k*2\symbol{94}(16)/18) then Prob2:=Prob2 +abs(NZ[y[k]+2])\symbol{94}2
  else Prob2:=Prob2 +abs(NZ[y[k]-2])\symbol{94}2 fi;  od:

$>$Prob2;

                    0.018781342656774252754

$>$Prob + Prob1 + Prob2 + abs(NZ[y[9]+2])\symbol{94}2; \# Yields probability that observed integer is in S-tilde; last term needed since both $y[9]+2$
and
$y[9]-2$ belong to S-tilde

0.89438284786571392115
\bigskip

{\bf Probability Calculation Using Formula (\ref{PtildeC})  for $\tilde{P}$}
\smallskip

$>$ PP:= (k,n,m,rp)$->$
2\symbol{94}k*2/(2\symbol{94}n*m)*sum(sum(evalf(sin(Pi*m*2\symbol{94}k*rp*(h+j/rp)/2\symbol{94}n)\symbol{94}2/

sin(Pi*2\symbol{94}k*rp*(h+ j/rp)/2\symbol{94}n)\symbol{94}2),j=1..floor(rp/2)),h=-2..1) + evalf((2\symbol{94}k
 -1)*m/2\symbol{94}n +
2*(2\symbol{94}k-1)/(2\symbol{94}n*m)*sum(sin(Pi*m*rp*2\symbol{94}k*w/2\symbol{94}n)\symbol{94}2/sin(Pi*rp*2\symbol{94}k*w/2\symbol{94}n)\symbol{94}2,w=1..2)):

$>$ PP(1,16,3641,9);

                     0.89438284786571368093 
\bigskip

\clearpage

\begin{flushleft}
{\Large \bf Appendix B: Proofs of Lower Bounds for Probability of Success}
\end{flushleft}
\bigskip

\begin{flushleft}
{\bf Lower Bound on $P$ Using Sine Series}
\end{flushleft}

Recall our formula (\ref{Pform}) for $P$:
$$
P = 2^{\k}\frac{2}{2^nm}\sum_{j=1}^{\floor{r'/2}}\frac{\sin^2\left(\frac{\pi
mj}{2^{n-\k}}\right)}{\sin^2\left(\frac{\pi j}{2^{n-\k}}\right)}+
(2^{\k}-1)\frac{m}{2^n},
$$
where we assume $r < N\le 2^{n_0}$, $2^\k r' = r$ with $r'\ge 3$  odd (and $\k\ge 0$), $\frac{2^n}{r} - 1 < m < \frac{2^n}{r} +1$, and 
$n > n_0$. Recall that $n_0$ is chosen to be the least positive integer such that $2^{n_0} \ge N$. Observe that $m > 2^n/r -1$ yields
\begin{equation}\label{seim}
(2^\k -1)\frac{m}{2^n} \ge \frac{1}{r'} - \frac{1}{r} - \frac{1}{2^{n-\k}} + \frac{1}{2^n} > \frac{1}{r'} - \frac{1}{r} - \frac{1}{2^{n-\k}}.
\end{equation} 
Also observe that if $j\in\left\{1, 2, \ldots, \floor{\frac{r'}{2}}\right\}$, then our inequalites for $m$ yield
\begin{equation}\label{cgss}
0 \le \frac{\pi m j}{2^{n-\k}} = \pi \frac{m r}{2^n} \left(\frac{j}{r'}\right) < \pi \left(1 + \frac{r}{2^n}\right)\left(\frac{1}{2}\right)
<\frac{\pi}{2}\left(1+ \frac{1}{2^{n-n_0}}\right)\le \frac{3\pi}{4}.
\end{equation}
  Our goal is to  establish the lower bounds (\ref{PLOD}) and (\ref{PLEVE}).  The work is tedious but straightforward.

 Using the sine function inequalities (\ref{sineEst}), the second of which holds by (\ref{cgss}),
as well as 
\begin{equation}
\label{SofS} (a)\ \ (1-x)^2 \ge 1- 2x \  {\rm for} \ x\in
(-\infty, \infty) \ \ {\rm and}\ \  (b)\ \ \sum_{j=1}^k j^2 = \frac{k(k+1)(2k+1)}{6}, 
\end{equation}
we have
\begin{eqnarray}
P &\ge& 2^{\k}\frac{2}{2^nm}\sum_{j=1}^{\floor{r'/2}}\frac{\left(\nr - \left(\nr\right)^3/6\right)^2}{(\dr)^2}+
(2^{\k}-1)\frac{m}{2^n}\ \ ({\rm by}\ (\ref{sineEst}))\nonumber\\
 &=& 2^{\k}\frac{2m}{2^n}\sum_{j=1}^{\floor{r'/2}}\left(1 - \left(\nr\right)^2/6\right)^2+
(2^{\k}-1)\frac{m}{2^n}\nonumber\\
 &\ge & 2^{\k}\frac{2m}{2^n}\sum_{j=1}^{\floor{r'/2}} \left(1 - \left(\nr\right)^2/3\right)+
(2^{\k}-1)\frac{m}{2^n}\ \ ({\rm by}\ (\ref{SofS}) (a))\nonumber\\
 & = & 2^{\k}\frac{2m}{2^n}\left(\floor{r'/2} - \left(\frac{\pi
m}{2^{n-\k}}\right)^2\ \frac13\sum_{j=1}^{\floor{r'/2}}j^2\right) + (2^{\k}-1)\frac{m}{2^n}\nonumber\\
 & = & 2^{\k}\frac{2m}{2^n}\floor{r'/2}\left(1 - \left(\frac{\pi
m}{2^{n-\k}}\right)^2\frac{(\floor{r'/2} + 1)(2\floor{r'/2} + 1)}{18}\right) +
(2^{\k}-1)\frac{m}{2^n} \ \ ({\rm by}\ (\ref{SofS}) (b))\label{GU}.
\end{eqnarray}
We continue the calculation, using (\ref{seim}), $\frac{2^n}{r} + 1> m > \frac{2^n}{r} - 1$, and  $\floor{r'/2} =
\frac{r'}{2} - \frac12$,
the latter fact holding because $r'$ is odd. We obtain separate underestimates for the cases 
\begin{itemize}
\item[(a)] $\k > 0$ (so that $r$, which equals $2^\k r'$, is even), and
\item[(b)]  $\k = 0$ (so that $r$ is odd and $r' = r$).
\end{itemize}

For $\k > 0$, we have
\begin{eqnarray}
P_{\even} & > &\left(\frac{2}{r'} - \frac{1}{2^{n-\k-1}}\right)\left(\frac{r'}{2} - \frac12\right)\left(1 -
\pi^2\left(\frac{1}{r'}+\frac{1}{2^{n-\k}}\right)^2\frac{(r' + 1)r'}{36}\right) + \frac{1}{r'}- \frac{1}{r} -
\frac{1}{2^{n-\k}}\nonumber\\
& = & \left(1- \frac{r}{2^n} - \frac{1}{r'} + \frac{1}{2^{n-\k}}\right)\left(1 -
\frac{\pi^2}{36}\left(\frac{r'+1}{r'} + \frac{r'+1}{2^{n-\k - 1}} + \frac{r'(r'+1)}{2^{2n-2\k}}\right)
\right)+ 
\frac{1}{r'}- \frac{1}{r} -
\frac{1}{2^{n-\k}}\label{TGEE}.
\end{eqnarray}
 
For case (b), note that when $\k = 0$ the final summand in (\ref{GU}) disappears.  Thus for $\k= 0$, so that $r = r'$, we have
\begin{eqnarray}
P_{\odd}& > &  \left(1- \frac{r}{2^n} - \frac{1}{r} + \frac{1}{2^{n}}\right)\left(1 -
\frac{\pi^2}{36}\left(\frac{r+1}{r} + \frac{r+1}{2^{n - 1}} + \frac{r(r+1)}{2^{2n}}\right)
\right)\nonumber\\
  & > &  \left(1- \frac{r}{2^n} - \frac{1}{r}\right)\left(1 -
\frac{\pi^2}{36}\left(\frac{r+1}{r} + \frac{r+1}{2^{n - 1}} + \frac{r(r+1)}{2^{2n}}\right)\label{SIFPO}
\right).
\end{eqnarray}
We analyze $P_\odd$ first. Recall that 
$r < N
\le 2^{n_0}$, where
$r$ is the order of
$b$ modulo
$N$.  For now, we just assume $n> n_0$. Note that if $2^{n_0}$ (or $2^{n_0} - 1$) is substituted into the quantity of (\ref{SIFPO}) for any $r$
appearing in the numerator of a fraction, the effect is to produce a smaller quantity;  thus, we have arrived at the advertised lower bound
(\ref{PLOD}) for
$P$ when
$r$ is odd:
$$
P_{\odd}  >  \left(1- \frac{1}{2^{n-n_0}} - \frac{1}{r}\right)\left(1 -
\frac{\pi^2}{36}\left(\frac{r+1}{r} + \frac{1}{2^{n -n_0 - 1}} + \frac{1}{2^{2(n-n_0)}}\right)
\right).
$$
We show that $P_{\odd} > .70$  assuming only that the difference $n-n_0 \ge 11$ and $r \ge 41$.  Thus if $N\ge
2^{11}$ and
$r\ge 40$ is odd, then Shor's algorithm, as it was described in his papers \cite{Shor1,Shor2}, finds a divisor of $r$ with
probability at least $70\%$.  Assume $n-n_0\ge 11$, then
$$
P_\odd > \left(1- \frac{1}{2^{11}} - \frac{1}{r}\right)\left(1 -
\frac{\pi^2}{36}\left(\frac{r+1}{r} + \frac{1}{2^{10}} + \frac{1}{2^{2(11)}}\right)
\right).
$$
Define $f:[41, \infty)\rightarrow \R$ by 
$$
f(r) =\left(1- \frac{1}{2^{11}} - \frac{1}{r}\right)\left(1 -
\frac{\pi^2}{36}\left(\frac{r+1}{r} + \frac{1}{2^{10}} + \frac{1}{2^{22}}\right)
\right).
$$
It is easy to show that $f$ has positive derivative on $[41, \infty)$ and $f(41) >  .70$, which verifies  our claims concerning
successfully finding a divisor of $r$ in case
$r$ is odd.  

Now we turn to the case $\k > 0$ so that  $r = 2^\k r'$ is even.  Using (\ref{TGEE}) along with $\k \le n_0$ and $r< 2^{n_0}$, we obtain the
advertised lower bound (\ref{PLEVE}) for $P$ given $r$  is even:
\begin{eqnarray*}
P_{\rm even} &> & \left(1- \frac{1}{2^{n-n_0}} - \frac{1}{r'}\right)\left(1 -
\frac{\pi^2}{36}\left(\frac{r'+1}{r'} + \frac{1}{2^{n-n_0-2}} +
\frac{1}{2^{2(n-n_0)-1}}\right)
\right)\\
& & \rule{2in}{0in} + \frac{1}{r'}- \frac{1}{2^\k r'} -
\frac{1}{2^{n-n_0}}.
\end{eqnarray*}  

We continue to assume that $r \ge 40$ and $n-n_0 \ge
11$.  Because $2^\k r' \ge 40$ we may work with the following four cases (1) $\k \ge 4, r' \ge 3$, (2) $\k = 3, r'\ge
5$, (3) $\k = 2, r'\ge 11$, and (4) $\k = 1, r'\ge 21$.   We handle these cases separately.
Case 1:  if we assume that $\k\ge 4$, we can say
$$
P_\even > \left(1- \frac{1}{2^{11}} - \frac{1}{r'}\right)\left(1 -
\frac{\pi^2}{36}\left(\frac{r'+1}{r'} + \frac{1}{2^{9}} + \frac{1}{2^{21}}\right)
\right)+ 
\frac{1}{r'}- \frac{1}{16r'} -
\frac{1}{2^{11}}.
$$
Define $f:[3, \infty)\rightarrow \R$ by 
$$
f(r') =\left(1- \frac{1}{2^{11}} - \frac{1}{r'}\right)\left(1 -
\frac{\pi^2}{36}\left(\frac{r'+1}{r'} + \frac{1}{2^{9}} + \frac{1}{2^{21}}\right)
\right)+ 
\frac{15}{16r'}-
\frac{1}{2^{11}}.
$$
It is easy to show that $f$ has a global minimum on $[3, \infty)$ at
$r_0:=\frac{4194304\pi^2}{9(524288- 569\pi^2)}\approx 8.87$ and that $f(r_0)> .72$.

Case 2: For $\k= 3$, $r'\ge 5$, we can say $P_\even > f(r'),
$
where $f:[5, \infty)\rightarrow \R$ is given by 
$$
f(r') =\left(1- \frac{1}{2^{11}} - \frac{1}{r'}\right)\left(1 -
\frac{\pi^2}{36}\left(\frac{r'+1}{r'} + \frac{1}{2^{9}} + \frac{1}{2^{21}}\right)
\right)+ 
\frac{7}{8r'}-
\frac{1}{2^{11}}.
$$
It is easy to show that $f$ has positive derivative on $[5, \infty)$ and $f(5) >  .71$.

Case 3: For $\k = 2$, $r'\ge 11$,  we can say
$
P_\even > f(r'),
$
where $f:[11, \infty)\rightarrow \R$ is given by 
 $$
f(r') =\left(1- \frac{1}{2^{11}} - \frac{1}{r'}\right)\left(1 -
\frac{\pi^2}{36}\left(\frac{r'+1}{r'} + \frac{1}{2^{9}} + \frac{1}{2^{21}}\right)
\right)+ 
\frac{3}{4r'}-
\frac{1}{2^{11}}.
$$
It is easy to show that $f$ has positive derivative on $[11,\infty)$ and $f(11) >  .70$.

Case 4:  For $\k= 1$, $r'\ge 21$, we can say
$
P_\even > f(r'),
$
where $f:[21, \infty)\rightarrow \R$ is given by 
$$
f(r') =\left(1- \frac{1}{2^{11}} - \frac{1}{r'}\right)\left(1 -
\frac{\pi^2}{36}\left(\frac{r'+1}{r'} + \frac{1}{2^{9}} + \frac{1}{2^{21}}\right)
\right)+ 
\frac{1}{2r'}-
\frac{1}{2^{11}}.
$$
It is easy to show that $f$ has positive derivative on $[21,\infty)$ and $f(21) >  .70$. 

The preceding four cases justify our claims concerning the probability $P$ of success when $r$ is even, and
thus, complete our proof that if Shor's algorithm is carried out with an input register having the size
described in Shor's original paper, then the probability of finding a divisor of the period sought exceeds
70\% (as long as $r\ge 40$ and $N\ge 2^{11}$).  
\bigskip

\begin{flushleft}{\bf Bounding $P$  Below by an Integral}
\end{flushleft}

 We provide a lower bound for $P$ in terms of an integral. We start with a representation of $P$ derived from our formula
(\ref{Pform}) and equation (\ref{FABY}):
\begin{equation}\label{PformU}
P=2^\k\frac{2}{2^nm}\sum_{j=1}^{\floor{r'/2}}\frac{\sin^2\left(\frac{\pi
m r (j/r')}{2^n}\right)}{\sin^2\left(\frac{\pi r (j/r')}{2^n}\right)}+
(2^{\k}-1)\frac{m}{2^n},
\end{equation}
where $r < N\le 2^{n_0}$, $\frac{2^n}{r} - 1 < m < \frac{2^n}{r} +1$, and $2^\k r' = r$ with $\k$ a nonnegative integer and $r'\ge 3$  odd.  We
assume that 
$n$ satisfies
$2^n\ge N^2$ and  that $r < N$.  Recall that since $r'$ is odd, $\floor{\frac{r'}{2}}  = (r'-1)/2$.

 Our approach to finding a integral-based lower bound for $P$ is
not the simplest possible one. We use methods here that will be required in our work to underestimate $\tilde{P}_q$ in the final subsection of
this appendix.

\begin{lemma}\label{sineuest} For $j\in \{1, 2, \ldots, \floor{\frac{r'}{2}}\}$, 
$$
\sin^2\left(\frac{\pi
m r (j/r')}{2^n}\right) \ge \sin^2\left(\frac{\pi j}{r'} - \frac{\pi
j}{2^{n-\k}}\right).
$$
\end{lemma}
\smallskip

Proof. Using $2^n/r+  1 \ge m \ge 2^n/r -1$, we see  that the argument of the
sine function  on the left in the lemma statement  satisfies
\begin{equation}\label{PDFLY}
\pi\left(\frac{j}{r'}+ \frac{j}{2^{n-\k}}\right) \ge \frac{\pi
m r (j/r')}{2^n}  \ge \pi\left(\frac{j}{r'} - \frac{j}{2^{n-\k}}\right).
\end{equation}
Note that the rightmost expression in (\ref{PDFLY}) is positive: $\pi j\left(\frac{1}{r'} - \frac{2^\k}{2^{n}}\right) =
\pi j\left(\frac{2^n-r}{2^{n}r'}\right) > 0$.  The following simple computation shows that leftmost quantity in (\ref{PDFLY})
 is less than $\pi/2$ for all $j$ between $1$ and
$\floor{r'/2} = (r'-1)/2$: 

Assuming $j\le \frac{r' -1}{2}$, we have
\begin{eqnarray*}
\pi\left(\frac{j}{r'}+ \frac{j}{2^{n-\k}}\right) &\le & \pi\left(\frac{1}{2} - \frac{1}{2r'}+
\frac{r'-1}{2\cdot2^{n-\k}}\right)\\
   & = & \frac{\pi}{2} - \frac{\pi}{2}\left(\frac{1}{r'} - \frac{2^\k(r'-1)}{2^{n}}\right)\\
 & = & \frac{\pi}{2} - \frac{\pi}{2}\left(\frac{2^{n} - r(r'-1)}{2^{n}r'}\right)\\
  & < & \frac{\pi}{2},
\end{eqnarray*} 	
where the inequality on the final line follows because the quantity inside parentheses on the penultimate line is positive ($2^n - r(r'-1)) > 2^n -
r^2 > 2^n- N^2\ge 0)$.  Thus, because the sine function is increasing  on $[0, \pi/2]$, we will obtain an underestimate 
of $\sin\left(\frac{\pi
m r (j/r')}{2^n}\right)$ by replacing 
$ \frac{\pi
m r (j/r')}{2^n}$ with  $\pi\left(\frac{j}{r'} - \frac{j}{2^{n-\k}}\right)$, which yields the lemma.///

\smallskip

\begin{lemma}\label{USLY} For real numbers $a$ and $b$ we have
$$
\sin^2(a \pm b) \ge (\sin^2 a)(1-b^2) - 2|b||\sin a|.
$$
\end{lemma}
\smallskip

Proof.  Using the angle addition formula for the sine function and then 
\begin{equation}\label{HTDWS}
(s - t)^2 \ge s^2 - 2st,
\end{equation} 
which is valid for all
real numbers $s$ and $t$, we find
\begin{eqnarray*}
\sin^2(a \pm b) &\ge& (\vstrut|\sin(a)\cos(b)| - |\sin(b)\cos(a)|)^2\\
                 & \ge & \sin^2(a)\cos^2(b) -2|\sin(b)||\sin(a)||\cos(a)\cos(b)|\\
         & \ge & \sin^2(a)(1-\sin^2(b)) - 2|\sin(b)||\sin(a)|\\
          & \ge & \sin^2(a)(1-b^2)-2|b||\sin(a)|.///
\end{eqnarray*}
\clearpage

 Using (\ref{PformU}), Lemma~\ref{sineuest}, and replacing 
$\sin^2\left(\frac{\pi r (j/r')}{2^n}\right)$ with the larger quantity $\left(\frac{\pi r (j/r')}{2^n}\right)^2$, we have
\begin{equation}\label{PFU}
P \ge  \left[\frac{2^\k(2)}{2^nm}\sum_{j=1}^{\floor{r'/2}}\frac{\sin^2\left(\frac{\pi j}{r'} - \frac{\pi
j}{2^{n-\k}}\right)}{\left(\frac{\pi r(j/r')}{2^n}\right)^2}\right]+
(2^{\k}-1)\frac{m}{2^n}.
\end{equation}
We seek to find an easily computable lower bound for the quantity in square brackets  in the preceding inequality; calling this
quantity
$Q$ we have
\begin{eqnarray}
Q& =& 
\frac{2^{\k+1}}{(\pi^2)\frac{mr}{2^n}}\left(\frac{1}{r}\sum_{j=1}^{\floor{r'/2}}\frac{\sin^2\left(\frac{\pi j}{r'} -
\frac{\pi j}{2^{n-\k}}\right)}{\left(\frac{j}{r'}\right)^2}\right)\label{Qstart} \\
&\ge &  \frac{2}{(\pi^2)\frac{mr}{2^n}}\left(\frac{1}{r'}\sum_{j=1}^{\floor{r'/2}}\frac{\sin^2\left(\frac{\pi j}{r'}\right)\left(1
-\left(\frac{\pi j}{2^{n-\k}}\right)^2\right) - \frac{\pi j}{2^{n-\k-1}}\sin\left(\frac{\pi
j}{r'}\right)}{\left(\frac{j}{r'}\right)^2}\right)\label{QEST}
\end{eqnarray} 
where to obtain (\ref{QEST}), we have used $\frac{2^\k}{r} = \frac{1}{r'}$ as well as Lemma~\ref{USLY} with $a = \frac{\pi j}{r'}$ and $b = \frac{\pi
j}{2^{n-\k}}$.  We continue the calculation, underestimating the quantity on line (\ref{QEST}) by replacing the first occurrence of
$\pi j/2^{n-\k}$ with $\frac{\pi r}{2^{n+1}}$, which exceeds its maximum possible value 
$\pi(r'-1)/2^{n-\k + 1}$,  replacing $\frac{mr}{2^n}$ with $(1 + \frac{r}{2^n})$, and separating the sum:
\begin{eqnarray}
Q  &\ge &\frac{2}{(\pi^2)\left(1 + \frac{r}{2^n}\right)}\left(\frac{1}{r'}\sum_{j=1}^{\floor{r'/2}}\frac{\sin^2\left(\frac{\pi
j}{r'}\right)\left(1-\left(\frac{\pi
r}{2^{n+1}}\right)^2\right)}{\left(\frac{j}{r'}\right)^2} - \frac{1}{r'}\sum_{j=1}^{\floor{r'/2}}\frac{\frac{\pi
j}{2^{n-\k-1}}\sin\left(\frac{\pi j}{r'}\right)}{\left(\frac{j}{r'}\right)^2}\right)\label{TBSQ}
\end{eqnarray}
We make the subtracted quantity in (\ref{TBSQ}) larger by  replacing $\sin(\pi j/r')$ with $\pi j/r'$; we also cancel $j$'s and $r'$'s,
obtaining
\begin{eqnarray}
Q &\ge &\frac{2\left(1-\left(\frac{\pi
r}{2^{n+1}}\right)^2\right)}{(\pi^2)\left(1 + \frac{r}{2^n}\right)}\left(\frac{1}{r'}\sum_{j=1}^{\floor{r'/2}}\frac{\sin^2\left(\frac{\pi
j}{r'}\right)}{\left(\frac{j}{r'}\right)^2}\right) - \frac{2}{(\pi^2)\left(1 + \frac{r}{2^n}\right)}\frac{\pi^2}{2^{n-\k - 1}}\floor{r'/2}\nonumber
\end{eqnarray}
We increase the subtracted quantity on the preceding line by replacing $1/(1 + r/2^n)$ with $1$ and we decrease  the initial quantity by
viewing the sum in parentheses as a Riemann sum with a left-endpoint selection for the decreasing function
$x\mapsto \sin^2(\pi x)/x^2$ on $[\frac{1}{r'}, \frac{1}{2} + \frac{1}{2r'}]$:
\begin{eqnarray}
Q &\ge & \frac{2\left(1-\left(\frac{\pi
r}{2^{n+1}}\right)^2\right)}{(\pi^2)\left(1 + \frac{r}{2^n}\right)}\int_{1/r'}^{\frac{1}{2} + \frac{1}{2r'}}\frac{\sin^2(\pi x)}{x^2}\, dx -
\frac{2^\k(r'-1)}{2^{n- 1}}\label{DIGT}\nonumber\\
&\ge &  \frac{1-\left(\frac{\pi r}{2^{n+1}}\right)^2}{1 + \frac{r}{2^n}}\left(\frac{2}{\pi^2}\int_{1/r'}^{\frac{1}{2} +
\frac{1}{2r'}}\frac{\sin^2(\pi x)}{x^2}\, dx\right) - \frac{r}{2^{n-1}}\label{QFEE}.
\end{eqnarray}
Thus, starting with
(\ref{PFU}) and using the definition of $Q$, the underestimate (\ref{QFEE}) for $Q$, as well as (\ref{seim}),
we have
$$
P \ge \frac{1-\left(\frac{\pi r}{2^{n+1}}\right)^2}{1 + \frac{r}{2^n}}\left(\frac{2}{\pi^2}\int_{1/r'}^{\frac{1}{2} +
\frac{1}{2r'}}\frac{\sin^2(\pi x)}{x^2}\, dx\right) -
\frac{r}{2^{n-1}}+\frac{1}{r'} - \frac{1}{r} -\frac{2^\k}{2^n}.
$$
Because $r< N\le 2^{n/2}$, we have $\frac{r}{2^n} \le \frac{1}{N}$; using this as well as $2^\k/2^n < r/2^n$ and $r = 2^\k r'$
yields
$$
P \ge \frac{1-\frac{\pi^2}{4N^2}}{1 + \frac{1}{N}}\left(\frac{2}{\pi^2}\int_{1/r'}^{\frac{1}{2} +
\frac{1}{2r'}}\frac{\sin^2(\pi x)}{x^2}\, dx\right) -
\frac{3}{N}+ \frac{1}{r'} -\frac{1}{2^\k r'},
$$
which is the advertised lower bound (\ref{PFinal}) on $P$.

\begin{flushleft}{\bf Bounding $\tilde{P}_q$  Below (Including $\tilde{P}_0 = \tilde{P}$)}
\end{flushleft}

 We derive the lower bound (\ref{GCPP}) for $\tilde{P}_q$, which upon letting $q=0$ yields the lower bound (\ref{NNPP}) for $\tilde{P}$.
We depend upon the results of the preceding subsection along with the following three Lemmas.

\begin{lemma}\label{hsumLem} $\sum_{h=1}^\infty \frac{1}{(h-\frac{1}{2})^2}\le 6$
\end{lemma}
\smallskip

Proof.  
 $$
\sum_{h=1}^\infty \frac{1}{(h -\frac12)^2} = \left(4 + \sum_{h=2}^\infty \frac{1}{(h
-\frac12)^2}\right) \le \left(4 + \int_{1/2}^\infty \frac{1}{x^2}\, dx\right) = 6./// 
$$

\begin{lemma}\label{GIEL}  For  every integer $h$ and every nonnegative integer $q$, 
$$
\sin^2\left(\frac{\pi m r (j/r' + h)}{2^{n+q}}\right) \ge\sin^2\left(\frac{\pi m r (j/r'))}{2^{n+q}}\right)\left(\vstrutt 1 -\left(\frac{\pi
hr}{2^{n+q}}\right)^2\right) - 2\left|\frac{\pi hr}{2^{n+q}}\right|.
$$
\end{lemma}

\smallskip

Proof.   Let $W = \sin^2\left(\frac{\pi m r (j/r' + h)}{2^{n+q}}\right)$.  We have
\begin{eqnarray*}
W&\ge & \left(\left|\sin\left(\frac{\pi m r (j/r'))}{2^{n+q}}\right)\cos\left(\frac{\pi h m r}{2^{n+q}}\right)\right| -
\left|\sin\left(\frac{\pi h m r}{2^{n+q}}\right)\cos\left(\frac{\pi m r (j/r'))}{2^{n+q}}\right)\right|\right)^2\\
&\ge & \sin^2\left(\frac{\pi m r (j/r'))}{2^{n+q}}\right)\cos^2\left(\frac{\pi h m r}{2^{n+q}}\right) - 2\left|\sin\left(\frac{\pi h m
r}{2^{n+q}}\right)\right|,
\end{eqnarray*}
where, to obtain the second inequality, we have used (\ref{HTDWS})  as well as 
$$
\left|\sin\left(\frac{\pi m r
(j/r'))}{2^{n+q}}\right)\cos\left(\frac{\pi h m r}{2^{n+q}}\right)\cos\left(\frac{\pi m r (j/r'))}{2^{n+q}}\right)\right|\le 1.
$$
We  continue the calculation, using $mr/2^{n+q} = 1 + x$, where $-r/2^{n+q} \le x \le r/2^{n+q}$:  
\begin{eqnarray*}
W&\ge & \sin^2\left(\frac{\pi m r (j/r'))}{2^{n+q}}\right)\left(\vstrutt 1 - \sin^2\left(\pi h (1+x)\right)\right) -
2\left|\vstrutt\sin\left(\pi h (1+x)\right)\right|\\
&\ge& \sin^2\left(\frac{\pi m r (j/r'))}{2^{n+q}}\right)\left(\vstrutt 1 - \cos^2(\pi h)\sin^2(\pi hx)\right) -
2\left|\vstrut\cos(\pi h)\sin(\pi hx)\right|\\
& \ge & \sin^2\left(\frac{\pi m r (j/r'))}{2^{n+q}}\right)\left(\vstrutt 1 -\left(\frac{\pi hr}{2^{n+q}}\right)^2\right) -
2\left|\frac{\pi hr}{2^{n+q}}\right|,
\end{eqnarray*}
as desired.///
\bigskip

\begin{lemma}  \label{KeyyLem}
For every nonzero integer $h$ and odd integer $r'\ge 3$,
\begin{equation}\label{neghc}
\frac{1}{r'}\sum_{j=1}^{\floor{r'/2}}\frac{\sin^2\left(\frac{\pi
j}{r'}\right)}{\left(\frac{j}{r'}+h\right)^2}  \ge \int_0^{\frac{1}{2}-\frac{1}{2r'}} \frac{\sin^2(\pi x)}{(x+h)^2}\, dx
\end{equation}
\end{lemma}
\smallskip

Proof.  For every integer $h$, let
\begin{equation}\label{fhif}
f_h(x) = \frac{\sin^2(\pi x)}{(x+h)^2}.
\end{equation}

Assume $h$ is a negative integer. As $x$ increases from $0$ to
$1/2$, $\sin^2(\pi x)$ increases and
$(x + h)^2$ decreases  (since 
$h
\le -1$). Thus $f_h$ is an increasing function of $x$ when $h$ is negative.  View the left-hand side of
(\ref{neghc}) as a Riemann sum for $f_h$ corresponding to the partition $\mathcal{P}:=\left\{[0, \frac{1}{r'}], [
\frac{1}{r'}, \frac{2}{r'}],
\ldots,
\left[\frac{\frac{r'}{2}-\frac{3}{2}}{r'}, \frac{\frac{r'}{2}-\frac{1}{2}}{r'}\right]\right\}$ of  $\left[0, \frac{1}{2} - \frac{1}{2r'}\right]$
with right-hand selection points $SP:=\{\frac{1}{r'}, \frac{2}{r'},
\ldots, \frac{\frac{r'}{2}-\frac{1}{2}}{r'}\}$.  Because,  $f_h$ is increasing on $\left[0, \frac{1}{2} -
\frac{1}{2r'}\right]$, the Riemann sum overestimates the integral.  Hence, (\ref{neghc}) holds.

 Now assume that $h$ is a positive integer.   In this case, the function $f_h$ 
increases up to a maximum occurring at ``$\x(h)$'', which is a little less than $1/2$, and then decreases. For example, $x_m(1)\approx 0.4303$ for the
function $f_1$ whose graph appears in Figure 2. To establish the lemma for positive $h$ we will need to used the following easily verified facts:
\begin{itemize}
\item[(a)] For every positive integer $h$, the point $\x(h)$ where $f_h$ attains its maximum value on $[0, 1/2]$ exceeds $0.43$.   
\item[(b)] For every positive integer $h$, there is a positive number $a< 1/4$ such that the graph of $f_h$ is concave up on $[0, a]$ and down on
$[a, 1/2]$.
\end{itemize}
 
Note that if $r' = 3$, $5$, or $7$, then the interval of integration on the  right of (\ref{neghc}) is contained in [0, 0.43]. Since $f_h$ is increasing
on $[0, 0.43]$ for every $h$,  (\ref{neghc})  holds for $r'=3, 5, 7$ by the argument applied above for negative values of $h$.   Thus we assume
$r'
\ge 9$.

For the remainder of the argument $j$ is used to denote
an integer in
$\{1, 2, \ldots, (r'-1)/2\}$. Define $j_a$ to be the least positive integer such that $\frac{j_a}{r'} - \frac{1}{2r'} > a$.   Because the graph of $f_h$ is concave down on $(a, 1/2]$, for all
$j
\geq j_a$ the integral
\begin{equation}
   \int_{\frac{j}{r'} - \frac{1}{2r'}}^{\frac{j}{r}'+\frac{1}{2r'}} f_h(x)\, dx
\end{equation}  is less than the area $f(j/r') 1/r'$ of the trapezoid (pictured in Figure 2) bounded by the $x$-axis, the vertical lines
$x=\frac{j}{r'} - \frac{1}{2r'}$, $x=\frac{j}{r'} + \frac{1}{2r'}$, and the line tangent to the graph of
$f_h$ at $(j/r', f(j/r'))$.
  It follows that 
\begin{equation} \label{int1}
\frac{1}{r'} \sum_{j=j_a}^{(r'-1)/2} f_h\left(\frac{j}{r'}\right) \geq \int_{\frac{j_a}{r'}-\frac{1}{2r'}}^{\frac{1}{2}} f_h(x)\, dx.
\end{equation}

\begin{center}

\includegraphics*[height=2.4in]{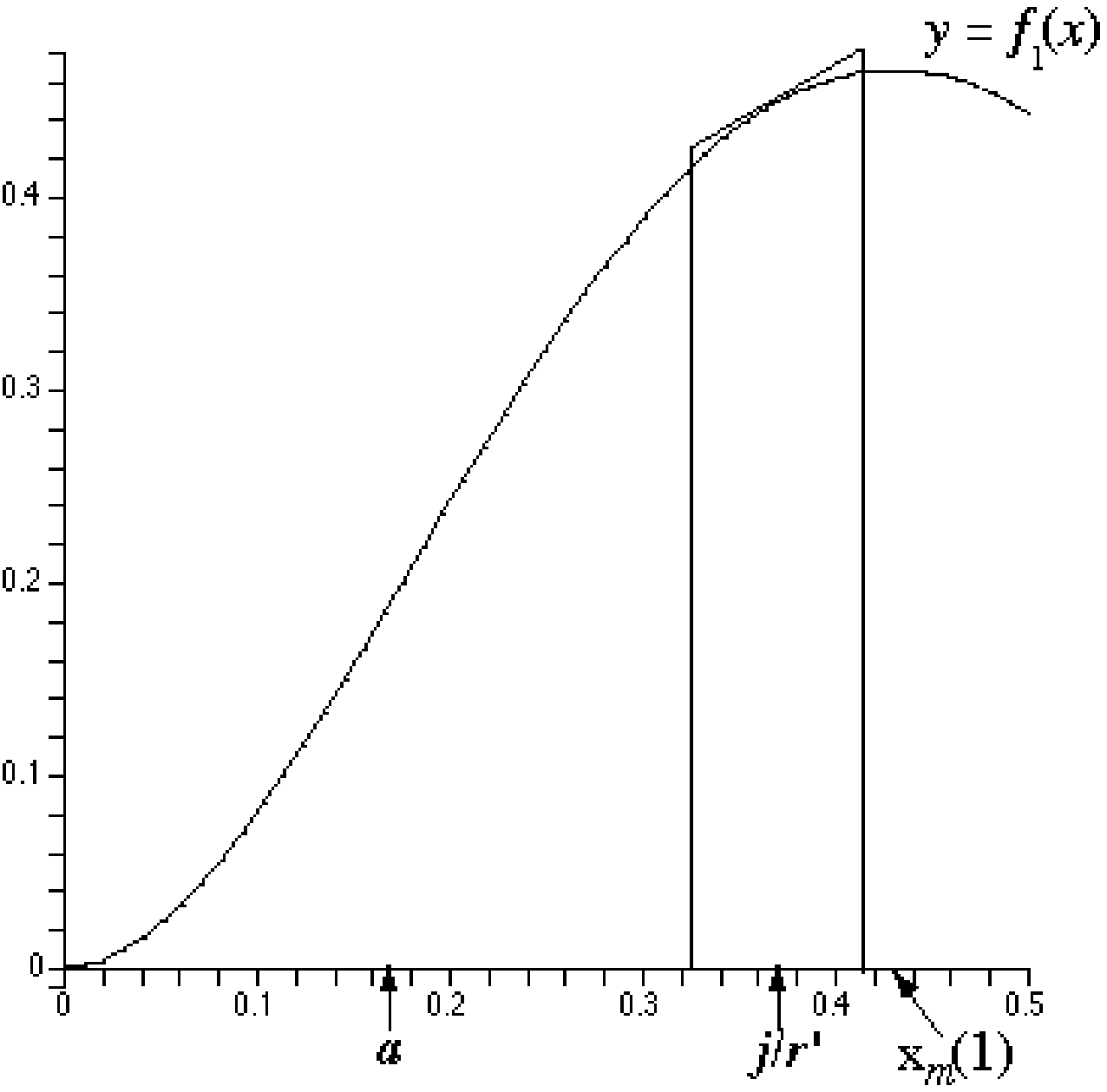}

\begin{minipage}{5in}
{\bf Figure 2}: The trapezoid pictured has area exceeding the integral of $f_1$ over $\left[\frac{j}{r'} - \frac{1}{2r'}, \frac{j}{r'} +
\frac{1}{2r'}\right]$. 
\end{minipage}

\end{center}

For values of $j < j_a$, note that $\frac{j}{r'} \le a + \frac{1}{2r'} \le \frac{1}{4} + \frac{1}{18} < 0.43$ so that $f_h$ is increasing on the
interval
$\left[\frac{j}{r'} - \frac{1}{r'}, \frac{j}{r'}\right]$. Hence for such a $j$, the integral
\begin{equation}
   \int_{\frac{j}{r'} - \frac{1}{r'}}^{\frac{j}{r'}} f_h(x)\, dx
\end{equation}  is less than $f(j/r') 1/r'$.   From this it follows that 
\begin{equation} \label{int2}
\frac{1}{r'} \sum_{j=1}^{j_a-1} f_h\left(\frac{j}{r'}\right) \geq \int_{0}^{\frac{j_a-1}{r'}} f_h(x)\, dx.
\end{equation}

Combining (\ref{int1}) and (\ref{int2}) yields
\begin{eqnarray}
\frac{1}{r'} \sum_{j=1}^{(r'-1)/2} f_h(\frac{j}{r'}) & > & \int_{0}^{\frac{j_a-1}{r'}} f_h(x)\, dx + \int_{\frac{j_a}{r'}-\frac{1}{2r'}}^{\frac{1}{2}}
f_h(x)\, dx\nonumber
\\
 & = & \int_{0}^{\frac{1}{2}-\frac{1}{2r'}} f_h(x)\, dx +\left(\int_{\frac{1}{2}-\frac{1}{2r'}}^{\frac{1}{2}} f_h(x)\, dx
-\int_{\frac{j_a-1}{r'}}^{\frac{j_a}{r'}-\frac{1}{2r'}} f_h(x)\, dx\right). \label{FEFH2}
\end{eqnarray}
We complete the proof of the lemma by establishing that the quantity in parentheses on the right
of (\ref{FEFH2}) is nonnegative. It suffices to show that the minimum value of $f_h$ on
$\left[\frac{1}{2}-\frac{1}{2r'},
\frac12\right]$ exceeds the maximum value of $f_h$ on $\left[\frac{j_a-1}{r'}, \frac{j_a}{r'} - \frac{1}{2r'} \right]$.  We continue to assume that $h$
is an arbitrary positive integer.  Since
\begin{equation}\label{TFDTP}
\frac{j_a}{r'} -\frac{1}{2r'} \le a + \frac{1}{r'} \le \frac14 + \frac19 < 0.43,
\end{equation}
 and
$f_h$ is increasing on $[0, .43]$, the maximum value of $f$ on $\left[\frac{j_a-1}{r'}, \frac{j_a}{r'} - \frac{1}{2r'} \right]$ is $f_h\left(
\frac{j_a}{r'} - \frac{1}{2r'}\right)$.    The minimum value of $f_h$ on
$\left[\frac{1}{2}-\frac{1}{2r'}, \frac12\right]$ occurs either at
$L:=\frac{1}{2}-\frac{1}{2r'}$ or at
$1/2$. A computation
shows $f_h(1/2) > f_h(1/4 + 1/9)$; thus by (\ref{TFDTP}) and the fact that $f_h$ is increasing on $[0, .43]$,  we may conclude $f_h(1/2)  > f_h\left(
\frac{j_a}{r'} - \frac{1}{2r'}\right)$, as desired.  As for $L$, there are two possibilities, (i) $L\in [x_m(h), 1/2]$ or (ii) $\frac{j_a}{r'}-
\frac{1}{2r'} < L < x_m(h)$.   In case (i), $f(L) > f_h(1/2) >f_h\left(
\frac{j_a}{r'} - \frac{1}{2r'}\right)$ and in case (ii), the desired inequality holds since $f_h$ is increasing on $[0, x_m(h)]$. ///
\bigskip

We are now in position to find a lower bound for $\tilde{P}_q$ in terms of integrals of the functions $f_h$ defined by (\ref{fhif}).  We begin
with our exact formula for $\tilde{P}_q$ from Section~\ref{PCLCSect}, obtaining an underestimate for $\tilde{P}_q$ by eliminating the final term in the
formula (which is clearly nonnegative):
\begin{equation}\label{PqC}
\tilde{P}_q \ge  2^{\k}\frac{2}{2^{n+q}m}\sum_{h=-2^{q+1}}^{2^{q+1}-1}\sum_{j=1}^{\floor{r'/2}}\frac{\sin^2\left(\frac{\pi
mr(j/r'+h)}{2^{n+q}}\right)}{\sin^2\left(\frac{\pi r (j/r'+h)}{2^{n+q}}\right)} +
(2^{\k}-1)\frac{m}{2^{n+q}}\label{TSWIN}.
\end{equation}
Using $r< N/2$, $N^2 \le 2^n$ and Lemma~\ref{GIEL}, we obtain
\begin{equation}\label{NTEY}
\sin^2\left(\frac{\pi
mr(j/r'+h)}{2^{n+q}}\right) \ge \sin^2\left(\frac{\pi m r (j/r'))}{2^{n+q}}\right)\left(\vstrutt 1 -\left(\frac{\pi h}{2^{q+1}N}\right)^2\right)
-
\frac{\pi |h|}{2^qN}.
\end{equation}
Because we are assuming
$j$ varies from $1$ to $\floor{r'/2}$, $\frac{j}{r'}\le\frac{1}{r'}\floor{\frac{r'}{2}} < 1/2$, and thus we have for every integer $h$,
\begin{equation}\label{htrb}
\frac{|h|}{\left(\frac{j}{r'} + h\right)^2} \le \frac{|h|}{\left(|h| - \frac{1}{2}\right)^2} \le 4.
\end{equation}

 Using (\ref{NTEY})  and (\ref{TSWIN}), we have
\begin{eqnarray}
\tilde{P}_q & \ge &2^{\k}\frac{2}{2^{n+q}m}\sum_{h=-2^{q+1}}^{2^{q+1}-1}\sum_{j=1}^{\floor{r'/2}}\frac{\sin^2\left(\frac{\pi m r
(j/r'))}{2^{n+q}}\right)\left(\vstrutt 1 -\left(\frac{\pi h}{2^{q+1}N}\right)^2\right) -
\frac{\pi |h|}{2^qN}}{\sin^2\left(\frac{\pi r (j/r'+h)}{2^{n+q}}\right)} +
(2^{\k}-1)\frac{m}{2^{n+q}}\nonumber\\
&\ge
&\frac{2^{\k+1}}{(\pi^2)\frac{mr}{2^{n+q}}}\sum_{h=-2^{q+1}}^{2^{q+1}-1}\left(\frac{1}{r}\sum_{j=1}^{\floor{r'/2}}\frac{\sin^2\left(\frac{\pi m
r (j/r'))}{2^{n+q}}\right)\left(\vstrut 1 -\left(\frac{\pi h}{2^{q+1}N}\right)^2\right) -
\frac{\pi |h|}{2^qN}}{ (j/r'+h)^2}\right) +
(2^{\k}-1)\frac{m}{2^{n+q}}\nonumber\\ 
& \ge & \frac{\left(\vstrut 1 -\left(\frac{\pi 2^{q+1}}{2^{q+1}N}\right)^2\right)
2^{\k+1}}{(\pi^2)\frac{mr}{2^{n+q}}}\sum_{h=-2^{q+1}}^{2^{q+1}-1}\left(\frac{1}{r}\sum_{j=1}^{\floor{r'/2}}\frac{\sin^2\left(\frac{\pi m r
(j/r')}{2^{n+q}}\right)}{ (j/r'+h)^2}\right) \label{CGIY}\\ 
 & &  \rule{1.6in}{0in}
-\frac{2}{(\pi^2)\frac{mr}{2^{n+q}}}\sum_{h=-2^{q+1}}^{2^{q+1}-1}\left(\frac{1}{r'}\sum_{j=1}^{\floor{r'/2}}\frac{\frac{\pi
|h| }{2^qN}}{ (j/r'+h)^2}\right)+ (2^{\k}-1)\frac{m}{2^{n+q}},\nonumber
\end{eqnarray}
where (\ref{CGIY}) follows from the line that precedes it by replacing the first occurrence of $h$ in the numerator with its maximum possible
absolute value (namely $2^{q+1}$), by separating the sum, and by using $\frac{2^\k}{r} = \frac{1}{r'}$. Continuing the calculation, we have

\begin{eqnarray}
\tilde{P}_q &\ge & \frac{\left(1 -
\left(\frac{\pi}{N}\right)^2\right)2^{\k+1}}{(\pi^2)\frac{mr}{2^{n+q}}}\sum_{h=-2^{q+1}}^{2^{q+1}-1}\left(\frac{1}{r}\sum_{j=1}^{\floor{r'/2}}\frac{\sin^2\left(\frac{\pi
m r (j/r')}{2^{n+q}}\right)}{ (j/r'+h)^2}\right)\nonumber \\ 
 & & \rule{1.0in}{0in}-\frac{2}{(\pi 2^q
N)\frac{mr}{2^{n+q}}}\sum_{h=-2^{q+1}}^{2^{q+1}-1}\left(\frac{1}{r'}\sum_{j=1}^{\floor{r'/2}}\frac{|h|}{ (|h|-1/2)^2}\right)+
(2^{\k}-1)\frac{m}{2^{n+q}}\nonumber\\
 & \ge & \left(1 -
\left(\frac{\pi}{N}\right)^2\right)\sum_{h=-2^{q+1}}^{2^{q+1}-1}\left[\frac{2^{\k+1}}{(\pi^2)\frac{mr}{2^{n+q}}}\left(\frac{1}{r}\sum_{j=1}^{\floor{r'/2}}\frac{\sin^2\left(\frac{\pi
m r (j/r')}{2^{n+q}}\right)}{ (j/r'+h)^2}\right)\right]\label{TSSYA}\\  & & \rule{1.4in}{0in}
-\frac{2^{q+2}8\floor{r'/2}}{(\pi 2^qNr')\frac{mr}{2^{n+q}}} + (2^{\k}-1)\frac{m}{2^{n+q}}\ \ ({\rm by}\ (\ref{htrb})).\nonumber
\end{eqnarray}
Using Lemma~\ref{sineuest}, with $n + q$ replacing $n$, we obtain, for $h\in \{-2^{q+1}, \ldots, 2^{q+1}-1\}$, a lower bound on the square-bracketed
quantity in (\ref{TSSYA}):
\begin{equation}\label{BTHFCB}
\frac{2^{\k+1}}{(\pi^2)\frac{mr}{2^{n+q}}}\left(\frac{1}{r}\sum_{j=1}^{\floor{r'/2}}\frac{\sin^2\left(\frac{\pi
m r (j/r')}{2^{n+q}}\right)}{ (j/r'+h)^2}\right) \ge \frac{2^{\k+1}}{(\pi^2)\frac{mr}{2^{n+q}}}
\left(\frac{1}{r}\sum_{j=1}^{\floor{r'/2}}\frac{\sin^2\left(\frac{\pi j}{r'} - \frac{\pi j}{2^{n+q-\k}}\right)}{ (j/r'+h)^2}\right).
\end{equation}
  The right-hand side of the preceding inequality, with $h = 0$, is identical to $Q$ of (\ref{Qstart}) with $n +q$ replacing $n$.  Thus our work
bounding $Q$ below culminating in (\ref{QFEE}) shows 
\begin{equation}\label{forhz}
\frac{2^{\k+1}}{(\pi^2)\frac{mr}{2^{n+q}}}\left(\frac{1}{r}\sum_{j=1}^{\floor{r'/2}}\frac{\sin^2\left(\frac{\pi
m r (j/r')}{2^{n+q}}\right)}{ (j/r'+0)^2}\right)\ge
 \frac{1-\left(\frac{\pi r}{2^{n+q+1}}\right)^2}{1 + \frac{r}{2^{n+q}}}\left(\frac{2}{\pi^2}\int_{1/r'}^{\frac{1}{2} +
\frac{1}{2r'}}\frac{\sin^2(\pi x)}{x^2}\, dx\right) - \frac{r}{2^{n+q-1}}.
\end{equation}
Thus we have a lower bound for the $h=0$ summand of (\ref{TSSYA}).   To bound below the other summands in (\ref{TSSYA}), i.e. those
corresponding to
$h\in
\{-2^{q+1},
\ldots, 2^{q+1}-1\}\setminus\{0\}$, we again cycle through the lower bound calculation for $Q$, (\ref{Qstart}) through (\ref{QFEE}); this time
with two substitutions: $n+q$ replacing $n$ and $j/r' + h$ replacing
$j/r'$ in the denominator.  Underestimate (\ref{TBSQ}) becomes 
\begin{equation}\label{GCTIU}
 \frac{2\left(1-\left(\frac{\pi
r}{2^{n+q+1}}\right)^2\right)}{(\pi^2)\left(1 + \frac{r}{2^{n+q}}\right)}\frac{1}{r'}\sum_{j=1}^{\floor{r'/2}}\frac{\sin^2\left(\frac{\pi
j}{r'}\right)}{\left(\frac{j}{r'} +h\right)^2} -  \frac{2}{(\pi^2)\left(1 +
\frac{r}{2^{n+q}}\right)}\frac{1}{r'}\sum_{j=1}^{\floor{r'/2}}\frac{\frac{\pi j}{2^{n+q-\k-1}}\sin\left(\frac{\pi
j}{r'}\right)}{\left(\frac{j}{r'}+h\right)^2}.
\end{equation}
  We make the subtracted quantity in
(\ref{GCTIU}) larger by replacing $(j/r' + h)^2$ with $(|h|-1/2)^2$, $\sin(\pi j/r')$ with $\pi j/r'$, and we also replace
$1/(1+\frac{r}{2^{n+q}})$
 with the larger number $1$. Thus the subtracted quantity in (\ref{GCTIU}) is less than 
\begin{eqnarray}
\frac{2}{(\pi^2)}\left(\frac{1}{r'}\sum_{j=1}^{\floor{r'/2}}\frac{\frac{\pi j}{2^{n+q-\k-1}}\left(\frac{\pi
j}{r'}\right)}{\left(|h|-\frac12\right)^2}\right) &\le &
\frac{2}{2^{n+q-\k-1}\left(|h|-\frac12\right)^2}\left(\frac{1}{r'^2}\sum_{j=1}^{\floor{r'/2}} j^2 \right).\nonumber
\end{eqnarray}
The quantity on the right in parentheses simplifies:
$$
\frac{1}{r'^2}\frac{\floor{\frac{r'}{2}}\left(\floor{\frac{r'}{2}} + 1\right)\left(2\floor{\frac{r'}{2}} +1\right)}{6} =
\frac{(r'-1)(r'+1)}{24r'}\le \frac{r' +1}{24},
$$
where we have used $\floor{\frac{r'}{2}} = \frac{r' -1}{2}$.  Thus the subtracted quantity (\ref{GCTIU}) is less than or equal to 
$$
   \frac{(r'+1)}{12\cdot2^{n+q-\k-1}\left(|h|-\frac12\right)^2}
$$
and, by Lemma~\ref{KeyyLem}, the initial quantity in (\ref{GCTIU}) is greater than or equal to 
$$
\frac{2\left(1-\left(\frac{\pi
r}{2^{n+q+1}}\right)^2\right)}{(\pi^2)\left(1 + \frac{r}{2^{n+q}}\right)}\int_0^{\frac{1}{2}-\frac{1}{2r'}} \frac{\sin^2(\pi x)}{(x+h)^2}\, dx.
$$
Using the preceding two observations  as well as $2^\k r' = r$, $r < \frac{N}{2} \le \frac{2^{n/2}}{2}$, and $r + 2^\k \le 2r$, we have 
\begin{eqnarray}
{\rm Quantity} (\ref{GCTIU}) &\ge& \frac{2\left(1- \left(\frac{\pi r}{2^{n+q +1}}\right)^2\right)}{(\pi^2)\left(1 +
\frac{r}{2^{n+q}}\right)}\int_0^{\frac{1}{2}-\frac{1}{2r'}} \frac{\sin^2(\pi x)}{(x+h)^2}\, dx - 
\frac{r + 2^\k}{12\cdot2^{n+q-1}\left(|h|-\frac12\right)^2}\nonumber\\
&\ge & \frac{1-\left(\frac{\pi}{2^{q+2}N}\right)^2}{1 +
\frac{1}{2^{q+1}N}}\left(\frac{2}{\pi^2}\int_0^{\frac{1}{2}-\frac{1}{2r'}} \frac{\sin^2(\pi x)}{(x+h)^2}\, dx\right) - \frac{1}{12\cdot N
2^{q-1}(|h|-1/2)^2}\nonumber\\
& = & C(q, N)\left(\frac{2}{\pi^2}\int_0^{\frac{1}{2}-\frac{1}{2r'}} \frac{\sin^2(\pi x)}{(x+h)^2}\, dx\right) - L(h)\label{umhne},
\end{eqnarray}
 where we have defined
\begin{equation}\label{CqnLD}
 C(q, N) = \frac{1-\left(\frac{\pi}{2^{q+2}N}\right)^2}{1 +
\frac{1}{2^{q+1}N}}\ \ {\rm and} \ \   L(h) = \frac{1}{12\cdot N
2^{q-1}(|h|-1/2)^2}\ \ {\rm for}\ \ h\ne 0.
\end{equation}
 Notice that for nonzero $h$, our
under-approximating integral from (\ref{umhne}) has limits from $0$ to $\frac12 - \frac{1}{2r'}$ whereas that for the $h=0$ case has limits from
$\frac{1}{r'}$ to
$\frac{1}{2} + \frac{1}{2r'}$.  To make our final lower-bound formula for
$\tilde{P}_q$ simpler, we adjust the under-approximating integral from (\ref{forhz}) for the $h = 0$ case as follows:
\begin{eqnarray*}
\int_{1/r'}^{\frac{1}{2} + \frac{1}{2r'}}\frac{\sin^2(\pi x)}{x^2}\, dx &= &\int_{0}^{\frac{1}{2} + \frac{1}{2r'}}\frac{\sin^2(\pi x)}{x^2}\,
dx- \int_0^{\frac{1}{r'}} \frac{\sin^2(\pi x)}{x^2}\, dx\\
 &> &\int_{0}^{\frac{1}{2} - \frac{1}{2r'}}\frac{\sin^2(\pi x)}{x^2}\,
dx- \frac{\pi^2}{r'},
\end{eqnarray*}
where we have used the nonnegativity of the integrand as well as $\sin^2(\pi x)/x^2\le \pi^2$ to obtain the inequality.  Thus (\ref{forhz})
becomes
\begin{eqnarray}
\frac{2^{\k+1}}{(\pi^2)\frac{mr}{2^{n+q}}}\left(\frac{1}{r}\sum_{j=1}^{\floor{r'/2}}\frac{\sin^2\left(\frac{\pi
m r (j/r')}{2^{n+q}}\right)}{ (j/r'+0)^2}\right) &\ge&
 \frac{1-\left(\frac{\pi r}{2^{n+q+1}}\right)^2}{1 + \frac{r}{2^{n+q}}}\left(\frac{2}{\pi^2}\int_{0}^{\frac{1}{2} -
\frac{1}{2r'}}\frac{\sin^2(\pi x)}{x^2}\, dx\right)\nonumber \\
  &  & \rule{1in}{0in} -\frac{1-\left(\frac{\pi r}{2^{n+q+1}}\right)^2}{1 + \frac{r}{2^{n+q}}}\left(\frac{2}{r'}\right) -
\frac{r}{2^{n+q-1}}\nonumber\\
 & \ge &  \frac{1-\left(\frac{\pi}{N2^{q+2}}\right)^2}{1 + \frac{1}{N2^{q+1}}}\left(\frac{2}{\pi^2}\int_{0}^{\frac{1}{2} -
\frac{1}{2r'}}\frac{\sin^2(\pi x)}{x^2}\, dx\right) - \frac{2}{r'} - \frac{1}{N2^q}\nonumber\\
 & = & C(q, N)\left(\frac{2}{\pi^2}\int_{0}^{\frac{1}{2} -
\frac{1}{2r'}}\frac{\sin^2(\pi x)}{x^2}\, dx\right) - L(0),\label{ttuzc}
\end{eqnarray}
where 
$
L(0) := \frac{2}{r'} + \frac{1}{N2^q}.
$

Using (\ref{ttuzc}) to bound below the $h=0$ term of the sum on line (\ref{TSSYA}) and using 
(\ref{umhne}) to bound below the terms corresponding to $h\ne 0$,  we obtain
\begin{eqnarray}
\tilde{P}_q &\ge&  \left(1 -
\left(\frac{\pi}{N}\right)^2\right)\sum_{h=-2^{q+1}}^{2^{q+1}-1} \left(C(q, N)\left(\frac{2}{\pi^2}\int_{0}^{\frac{1}{2} -
\frac{1}{2r'}}\frac{\sin^2(\pi x)}{(x+h)^2}\, dx\right)  - L(h)\right)\label{NTLSYea}\\
& &\rule{1.5in}{0in} -\frac{2^{q+2}8\floor{r'/2}}{(\pi 2^qNr')\frac{mr}{2^{n+q}}} + (2^{\k}-1)\frac{m}{2^{n+q}}.\nonumber
\end{eqnarray}
The preceding inequality will yield the advertised bound (\ref{GCPP}) for $\tilde{P_q}$ after a few more steps.
Using $\frac{1}{r'}\floor{r'/2} < 1/2$, $m\ge \frac{2^{n+q}}{r}-1$, and $r< N/2$, we have
\begin{equation}\label{ILT1}
\frac{2^{q+2}8\floor{r'/2}}{(\pi 2^qNr')\frac{mr}{2^{n+q}}}\le \frac{16}{\pi N\left(1-
\frac{1}{N2^{q+1}}\right)}.
\end{equation} 
Using $2^\k < r < N/2$ and again using $m \ge \frac{2^{n+q}}{r}-1$, we get
\begin{equation}\label{ILT2}
(2^{\k}-1)\frac{m}{2^{n+q}} \ge \frac{1}{r'} - \frac{1}{r}  -\frac{2^\k}{2^{n+q}} + \frac{1}{2^{n+q}} >  \frac{1}{r'} - \frac{1}{2^\k r'} 
-\frac{1}{N2^{q+1}}.
\end{equation}
Finally,
\begin{eqnarray}
\sum_{h=-2^{q+1}}^{2^{q+1}-1} L(h) &=&  L(0) + \sum_{\begin{array}{c}h=-2^{q+1}\\h\ne 0\end{array}}^{2^{q+1}-1}
\frac{1}{12\cdot N 2^{q-1}(|h|-1/2)^2}\nonumber\ \ (\rm by\ (\ref{CqnLD}))\\
 & \le  &  \frac{2}{r'} + \frac{1}{N2^q} + \frac{1}{12\cdot N
2^{q-1}}\left(2\sum_{h=1}^\infty\frac{1}{(|h|-1/2)^2} \right)\nonumber\\
&\le &  \frac{2}{r'} + \frac{1}{N2^q} + \frac{1}{N
2^{q-1}} =  \frac{2}{r'} + \frac{3}{N2^q}\nonumber,
\end{eqnarray}
where Lemma~\ref{hsumLem} provides the final inequality.

Beginning with (\ref{NTLSYea}) and then using (\ref{ILT1}), (\ref{ILT2}), and $\sum_{h=-2^{q+1}}^{2^{q+1}-1} L(h) =  \frac{2}{r'} + \frac{3}{N2^q}$, we
have
\begin{eqnarray*}
\tilde{P}_q &\ge&  \left(1 -
\left(\frac{\pi}{N}\right)^2\right)C(q,N)\sum_{h=-2^{q+1}}^{2^{q+1}-1}\left( \frac{2}{\pi^2}\int_{0}^{\frac{1}{2} -
\frac{1}{2r'}}\frac{\sin^2(\pi x)}{(x+h)^2}\, dx\right)  -  \left(1 -
\left(\frac{\pi}{N}\right)^2\right)\left(\frac{2}{r'} + \frac{3}{N2^q}\right)\\
& &\rule{1.5in}{0in} -\frac{16}{\pi N\left(1-
\frac{1}{N2^{q+1}}\right)} + \frac{1}{r'} - \frac{1}{2^\k r'} 
-\frac{1}{N2^{q+1}}.
\end{eqnarray*}
Substituting $1$ for  $\left(1 -
\left(\frac{\pi}{N}\right)^2\right)$ the second time it appears on the right of the preceding inequality, simplifying, and using the definition
of $C(q,N)$, we arrive at the advertised lower bound on $\tilde{P}_q$:
\begin{eqnarray*}
\tilde{P}_q &\ge&  \frac{\left(1 -
\left(\frac{\pi}{N}\right)^2\right)\left(1-\left(\frac{\pi}{2^{q+2}N}\right)^2\right)}{1 +
\frac{1}{2^{q+1}N}}\sum_{h=-2^{q+1}}^{2^{q+1}-1}\left( \frac{2}{\pi^2}\int_{0}^{\frac{1}{2} -
\frac{1}{2r'}}\frac{\sin^2(\pi x)}{(x+h)^2}\, dx\right)  - \frac{1}{r'}\\
& &\rule{2.4in}{0in}  - \frac{7}{N2^{q+1}} -\frac{16}{\pi N\left(1-
\frac{1}{N2^{q+1}}\right)}  - \frac{1}{2^\k r'}.
\end{eqnarray*}


\begin{thebibliography}{9}
\bibitem{EkJo} A.\ Ekert and R.\ Josza, Quantum computation and Shor's factoring algorithm, {\it Rev.\ Mod.\ Phys.\/} 68 (1996), 733--753.
\bibitem{Ger} E.\ Gerjuoy, Shor's factoring algorithm and modern
cryptography.  An illustration of the capabilities inherent in quantum
computers., {\it Am.\ J.\ Phys.\/} 73 (2005), 521--540.
\bibitem{Her} I.\ N.\ Herstein, {\it Topics in Algebra}, Wiley, New York, 1975.
  
\bibitem{IR} I.\ Ireland and M.\ Rosen, {\it A Classical Introduction to Modern
Number Theory}, Springer-Verlag, New York, 1990

\bibitem{Mer} D.\ Mermin, {\it Lecture Notes on Quantum Computation},
Cambridge University Press, to appear.  (Draft available at
http://people.ccmr.cornell.edu/~mermin/qcomp/CS483.html)

\bibitem{NC} M.\ A. Nielsen and I.\ L.\ Chuang, {\it Quantum Computing and
Quantum Information}, Cambridge University Press, Cambridge 2000.

\bibitem{Hir} M.\ Hirvensalo, {\it Quantum Computing}, Springer, New York,
2001.

\bibitem{Shor1} P.\ Shor, Algorithms for quantum computation: discrete
logarithms and factoring, {\it Proc. 35nd Annual Symposium on Foundations
of Computer Science} (Shafi Goldwasser, ed.), IEEE Computer Society Press
(1994), 124-134.

\bibitem{Shor2} P.\ Shor, Polynomial time algorithms for Prime
Factorization and discrete logarithms on a quantum computer, {\it SIAM J.
Computing} 26, pp. 1484-1509 (1997).
\end{thebibliography}
\end{document}